\documentclass[preprint2]{aastex}

\newcommand{\myemailFF}{Fabio.Frescura@wits.ac.za}
\newcommand{\myemailCE}{chrise@uj.ac.za}

\shorttitle{Significance of Periodogram Peaks} \shortauthors{Frescura, Engelbrecht
and Frank}

\begin{document}


\title{Significance Tests for Periodogram Peaks}


\author{F. A. M. Frescura}
\affil{Centre for Theoretical Physics, University of the Witwatersrand, Private Bag
3, WITS 2050, South Africa} \email{\myemailFF}

\author{C. A. Engelbrecht}
\affil{Department of Physics, University of Johannesburg, PO Box 524, AUCKLAND PARK
2006, South Africa} \email{\myemailCE}

\and

\author{B. S. Frank}
\affil{School of Physics, University of the Witwatersrand, Private Bag 3, WITS 2050,
South Africa}




\begin{abstract}
We discuss methods currently in use for determining the significance of peaks in the
periodograms of time series. We discuss some general methods for constructing
significance tests, false alarm probability functions, and the role played in these
by independent random variables and by empirical and theoretical cumulative
distribution functions. We also discuss the concept of ``independent frequencies" in
periodogram analysis. We propose a practical method for estimating the significance
of periodogram peaks, applicable to all time series irrespective of the spacing of
the data. This method, based on Monte Carlo simulations, produces significance tests
that are tailor-made for any given astronomical time series.
\end{abstract}


\keywords{Methods: data analysis --- Methods: statistical --- Stars: oscillations }

\section{INTRODUCTION}

Periodogram analysis is a vital ingredient of asteroseismology. It is used to
identify periodicities of oscillations in the observed star. Typically, the data
analysed are noisy. The effect of noise in the data is to produce spurious peaks in
the periodogram which arise, not because of any periodicity in the observed system,
but because of the way that the noisy signal has been sampled. These spurious peaks
can be surprisingly large. It is essential therefore to have reliable tests by which
to determine the significance of periodogram peaks.

This topic has already received attention in the literature. Key classical papers
include those of \citet{deeming75}, \citet{lomb76}, \citet{scargle82}, and
\citet{hb86}. We discuss pertinent aspects of these papers in the sections that
follow. More recently, criticisms of these papers have appeared in the work of
\citet{koen90} and Schwarzenberg-Czerny (1998), amongst others. Much of the
criticism has revolved around the appropriate means for attaching significance to
peaks that arise in a calculated periodogram.

Significance tests for periodograms are hugely important to the asteroseismologist
who relies on periodograms to deliver precise values for purported eigenfrequencies
of pulsation. Comparison of the values of observationally determined
eigenfrequencies with the values predicted by the latest theoretical models should,
in principle, allow the identification of modes actually excited in real stars, and,
subsequently, allow for asteroseismological analysis of those stars.

Asteroseismology appears to be on the threshold of a golden age, as extensive
surveys like ASAS \citep{pojmanski98}, and space missions in the mold of COROT
\citep{baglin02}, hugely increase the number of known pulsating stars, as well as
the time coverage available for their analysis. It is expected that periodogram
analysis will continue to play a prominent role in asteroseismology. Hence, accurate
interpretation of periodogram peaks is an issue of prime importance.

In this paper, we consider methods currently used for assessing the relevance of
periodogram peaks, and propose a practical method applicable to all time series,
irrespective of the spacing of the data. This method produces significance tests
that are tailor-made for any given astronomical time-series.

The structure of this paper is as follows. We first discuss the construction of
significance tests in general, and of Scargle's significance test in particular. We
consider the concept of ``independent frequencies" in periodogram analysis. We
comment on aspects of the work reported by \citet{scargle82}, \citet{hb86}, and
\citet{sc98}. We report our attempts at reproducing the results of \citet{hb86} by
Monte Carlo simulation, and discuss our failure to reproduce their results in
detail. The conclusions forced on us by the discrepancies between our results and
theirs lead us to the main points made in this paper. They also lead us to propose a
pragmatic method, applicable to all time series, for assessing the significance of
periodogram peaks. En route, we also discuss the problem of over-sampling the
periodogram.

Definitions of the periodogram assumed in our discussion are given in Appendices
\ref{classicalP} and \ref{scargleP} of this paper. Detailed discussions of the
phenomena of aliasing and spectral leakage, to which we refer in the text, may be
found in \citet{deeming75} and \cite{scargle82}.

\section{SIGNIFICANCE TESTS}

Noisy data produce noisy periodograms. Peaks in a periodogram may therefore not be
due to the presence of any real periodic phenomenon at all. They may simply be
random fluctuations in periodogram power caused by the presence of a noise component
in the data. Peaks arising in this way are spurious: they are not due to any real
periodicity in the observed phenomena, but are simply artifacts of chance events in
the accompanying noise.

Simulations show that noise in a time series can produce surprisingly large spurious
peaks in the associated periodogram. It is important therefore to develop reliable
tests for determining whether a given periodogram peak reflects a real periodicity
in the data, or is simply an artifact of the noise. In this section, we consider the
theoretical basis for a class of general, model-independent, tests. These determine
the probability that the periodogram powers observed in a data set might have arisen
from pure noise alone, with no other form of signal present. For a definition or
pure noise, see Appendix \ref{noiseApp}.

Note that, in this paper, we do not use the word ``power" in the formal statistical
sense, where it means the probability of rejection of the null hypothesis given that
the null hypothesis is false, but in its accepted physical sense. Thus ``periodogram
power" at frequency $\omega$ means $P_X(\omega)$, as defined in Appendices
\ref{classicalP} and \ref{scargleP}.

The basis for this general class of tests is the cumulative distribution function
(CDF),
\begin{eqnarray}
    F_Z(z) = {\rm Pr} [ Z \leq z ]
    \label{st-1}
\end{eqnarray}
where, the random variable $Z=P_X(\omega)$ is the periodogram power at frequency
$\omega$ for the time series $X$, and $z$ is some selected power threshold. The
function $F_Z(z)$ gives the probability that, when the data $X$ are pure noise,
their periodogram power at the given frequency $\omega$ does not rise above
power-level $z$.

Suppose a model of the observed system predicts an oscillation at frequency
$\omega$. Then, we expect $P_X(\omega)$ to be large at this frequency. However, pure
noise by itself might equally well produce a large value of $P_X(\omega)$. The
CDF in equation (\ref{st-1}) provides an objective criterion by which to determine
whether the observed large value of $P_X(\omega)$ is due to the presence of a {\em
bona fide} signal, or is nothing more than a spurious large fluctuation due only to
the presence of noise. Suppose the data $X$ are pure noise. Then the probability
that the normalised periodogram power at this frequency is less than a specified
value $z_0$ is
\begin{eqnarray}
    p_0 = F_Z(z_0)
\end{eqnarray}
Inverting this function,
\begin{eqnarray}
   z_0 =  {F_Z}^{-1}(p_0)
\end{eqnarray}
we obtain, for given $p_0$, the threshold power-level $z_0$ for which a power value
$Z\leq z_0$ has a probability $p_0$ of being due to pure noise alone. Equivalently,
a power value at frequency $\omega$ that exceeds $z_0$ has probability $1-p_0$ of
being due to pure noise alone. This test is both primitive and negative. It does
{\em not} tell us that $p_0$ is the probability that our signal contains a periodic
component of frequency $\omega$, but only that $p_0$ is the probability that our
signal is {\em not pure noise}.

In practice, one does not evaluate the periodogram power at a single frequency only,
but at a selected set $\{ \omega_\mu : \mu = 1, 2, ..., N \}$ of frequencies. The
procedure normally followed when looking for periodicities in data is this: the
periodogram is evaluated at the selected frequencies $\omega_\mu$, and the
periodogram power $P_X(\omega_\mu)$ is plotted against $\omega_\mu$; this plot is
then scanned for its highest peaks. The conclusion one would like to draw from the
plot is that peaks that rise substantially above all others indicate the presence of
genuine periodicities in the observed system. However, before we can have confidence
in this conclusion, we need first to rule out the possibility that the observed
periodogram plot could have been produced by pure noise alone. This is done by
calculating the probability that {\em the entire observed periodogram profile} could
be produced by pure noise alone. Suppose $X$ is pure noise. Consider the probability
that {\em all} of the periodogram powers $\{P_X(\omega_\mu) : \mu = 1,2,...,N\}$ at
the sampled frequencies $\{\omega_\mu\}$ fall below a specified power threshold $z$.
Define a new random variable,
\begin{eqnarray}
    Z_{\rm max} = {\rm sup}\ \{ P_X(\omega_\mu) : \mu=1,2,...,N\}
\end{eqnarray}
Thus $Z_{\rm max}$ is the maximum periodogram power among the set of $N$ {\em
sampled} powers. Now, the power at {\em each} of the sampled values will fall below
some specified threshold $z$ if and only if $Z_{\rm max}\leq z$. We thus need to
calculate the CDF
\begin{eqnarray}
  F_{Z_{\rm max}} (z) = {\rm Pr} [ Z_{\rm max} \leq z ]
\end{eqnarray}
The function $F_{Z_{\rm max}}(z)$ gives the probability that, when the data $X$ are
pure noise, the periodogram power $P_X(\omega_\mu)$ does not rise above the
threshold $z$ at {\em any} of the sampled frequencies $\{ \omega_\mu \}$. We
construct the second significance test as follows. Let $z_0$ be a specified power
threshold. The probability that pure noise alone will produce periodogram powers
$P_X(\omega_\mu)$ that do {\em not} exceed the threshold $z_0$ at {\em any} of the
sampled frequencies $\{ \omega_\mu \}$ is given by
\begin{eqnarray}
  p_0 = F_{Z_{\rm max}} (z_0)
\end{eqnarray}
Inverting this function,
\begin{eqnarray}
   z_0 =  {F_{Z_{\rm max}}}^{-1}(p_0)
\end{eqnarray}
For given $p_0$, this inverse function defines a threshold power-level $z_0$ such
that, if the periodogram power at each of the frequencies $\{\omega_\mu \}$ has
value $Z\leq z_0$, then the observed periodogram profile has probability $p_0$ of
being due to pure noise alone. This test reduces the probability of spurious
detections.


\section{SCARGLE'S SIGNIFICANCE TEST}

If the data are Gaussian pure noise, the periodogram power $Z=P_X(\omega)$ at any
given frequency $\omega$ of the sampled signal $X_k$ is exponentially distributed
with probability density function defined by (Scargle 1982, p 848),
\begin{eqnarray}
  \lefteqn{ p_Z (z)\ dz = {\rm Pr} [ z < Z < z+dz ] } \nonumber  \\
  && = \frac{1}{\sigma_X^2} \ e^{-z/  \sigma_X^2}\ dz
\end{eqnarray}
The cumulative distribution function is thus given by
\begin{eqnarray}
  \lefteqn{ P_Z (z) = {\rm Pr} [ Z < z ]  } \nonumber \\
  && = \int_{\zeta=0}^z \ p_Z(\zeta)\ d\zeta
  = 1-  e^{-z/ \sigma_X^2}
\end{eqnarray}
We are interested in the probability that the periodogram power at the given
frequency is greater than a specified threshold $z$. This is given by
\begin{eqnarray}
  {\rm Pr} [ Z > z ] = 1- P_Z (z) = e^{-z/ \sigma_X^2}
\end{eqnarray}
As the observed power $z$ becomes larger, it becomes exponentially less likely that
so high a power level (or higher) could be produced by pure noise alone, and
correspondingly more likely that the observed power level is due to a genuine
deterministic (i.e., non-noise) feature in the measured signal. Of course, this does
not mean necessarily that the suspected deterministic signal is {\em harmonic} with
frequency $\omega$, but simply that it is unlikely that this high power is due to
the noise component alone.

It is worth noting that the argument of the exponential in the cumulative
distribution function is not simply the observed power $z$, but the ratio
$z/\sigma^2_X$, which is the ratio of the periodogram power to the total variance of
the data (called total input signal power by some). This is an important point,
worth emphasising, as did \citet{hb86}. If the incorrect power ratio is used, then
the statistical tests considered by Scargle will necessarily fail. Thus,
normalisation of the periodogram power by the number $N_0$ of data points used to
calculate the periodogram (classical normalisation), or by the residual power after
a sine curve has been removed from the data, or by the variance of the observational
uncertainty, all lead to completely different statistical distributions for the
periodogram power and invalidate Scargle's analysis summarised in this paper. Of
course, this does not make alternative normalisations ``wrong". It does mean however
that they must be accompanied by alternative statistical analyses \citep{sc98}.

In practice, we do not evaluate the periodogram power at a single frequency alone,
but at a set of conveniently chosen frequencies $\{\omega_\mu : \mu = 1,2,...,N \}$.
We shall return to the question of how to choose these frequencies in a later
section. For the moment, suppose that we have the values of $P_X$ not at one value
of the frequency alone, but over a set of frequencies. This enables us to devise a
stronger test in which we determine the probability that the observed periodogram
power {\em over the entire set} of sampled frequencies have been produced by pure
noise alone.

To develop this new, stronger statistical test, we need to assume with Scargle that
we have evaluated the periodogram power at a set $\{\omega_\mu : \mu = 1,2,...,N_i
\}$ of frequencies chosen in such a way that the random variables $\{ Z_\mu =
P_X(\omega_\mu) : \mu = 1,2,...,N_i \}$ are {\em mutually independent}. \citet{hb86}
refer to a set of frequencies chosen in this way as ``independent frequencies". This
is an abuse of terminology, since it is not the frequencies that are ``independent",
but the random variables $Z_\mu$. However, this lack of precision leads to no
ambiguity and so is tolerable.

A large body of theorems is available for use if the random variables under
consideration are independent. Abandoning the condition of independence creates
serious complications in both the reasoning and the proofs of the results.

Suppose we observe a periodogram power at one of the $\omega_\mu$, that is higher
than a given threshold $z$. We ask, what is the probability that pure noise alone
could have produced a periodogram power of this level or higher among all of the
sampled independent periodogram frequencies? First, we calculate the probability
that {\em all} the sampled periodogram powers are less than the threshold power $z$.
Define
 \begin{eqnarray*}
  Z_{\rm max}= {\rm sup }\ \{ Z_1, Z_2, ..., Z_{N_i} \}
  \end{eqnarray*}
The probability that any given power $Z_\mu$ in this set falls below the threshold
is
 \begin{eqnarray*}
  {\rm Pr\ }[ Z_\mu < z ] = 1 - e^{-z/\sigma_X^2}
  \end{eqnarray*}
Since the $Z_\mu$ are independent, the probability that they all fall below the
threshold $z$ is given by
 \begin{eqnarray*}
   \lefteqn{ {\rm Pr\ }[ Z_1 < z \mbox{ and } Z_2 < z \mbox{ and }...
   \mbox{ and } Z_{N_i} < z
  ]  } \hspace*{1cm} \\
  &=&  {\rm Pr\ }[ Z_1 < z ] \ {\rm Pr\ }[ Z_2 < z ] \ ...\ {\rm Pr\ }[ Z_{N_i} < z
  ] \\
  &=& \left[ 1 - e^{- z/\sigma_X^2} \right]^{N_i}
  \end{eqnarray*}
The probability that {\em not all} the powers $Z_\mu$ are less than the threshold
$z$, that is, the probability that {\em at least one} of the powers $Z_\mu$ is above
the threshold $z$, is then,
 \begin{eqnarray}
 {\rm Pr\ }[ Z_{\rm max} > z ]  &=& 1 - \left[ 1 - e^{- z/\sigma_X^2} \right]^{N_i}
  \label{scargle-14}
  \end{eqnarray}
This is the function that Scargle proposes as a false alarm probability. The idea is
that we choose a probability, say $p_A$, that we regard as an acceptable level of
risk for the false detection of real deterministic signals. We solve the above
formula for $z$, to get a reference power threshold level $z_A$ given by
 \begin{eqnarray}
  z_A &=& - \sigma_X^2 \ln \left[ 1- \left( 1 - p_A \right)^{1/N_i} \right]
  \end{eqnarray}
Then, if we claim a detection whenever the power level at one of the frequencies
$\{\omega_\mu : \mu = 1,2,...,N_i \}$ exceeds the reference level $z_A$, the
probability that we will be {\em wrong} is given by $p_A$. That is, on average we
will be wrong only $p_A$ of the time, since pure noise can produce fluctuations
above this level at these frequencies only $p_A$ of the time.


\section{INDEPENDENT FREQUENCIES}

Scargle's test is constructed on the assumption that we can identify a set of
frequencies at which the periodogram powers are independent random variables. In the
case where the time-domain data are evenly spaced, we are guaranteed the existence
of such a set. These are called the {\em natural frequencies} \citep{scargle82}, or
the {\em standard frequencies} \citep{priestley81}. These are given by
 \begin{eqnarray}
  \omega_k  &=& \frac{2\pi k}{T}
  \end{eqnarray}
where $T$ is the total time span of the data set, that is, $T=t_{N_0}-t_1$, and $k=
0,...,[N_0/2]$, where $[N_0/2]$ signifies the integer part of $N_0/2$. The
statistics of $P_X(\omega_k)$ with $k=0$ are different from those with $k\neq 0$
\citep{priestley81}. If we omit $P_X(\omega_0)$, this leaves us with at most
$[N_0/2]$ independent frequencies. In practice, the omission of $\omega_0$ from the
set of independent frequencies is of no consequence. This frequency corresponds to a
DC component in the signal which is generally removed from the data before their
periodogram is calculated. Thus, in the case of evenly spaced data, we can easily
construct the Scargle false alarm probability function and apply it to determine the
significance of high periodogram-power levels at these ``independent frequencies".

It is worth emphasising that, since the false alarm probability function assumes
independent powers at the examined frequencies, {\em we can only use it to put a
significance level on the values of the periodogram-power at the chosen independent
frequencies}. Peaks found at other frequency values by over-sampling the periodogram
{\em cannot} be assessed in this way.

In the unevenly sampled case, the situation changes dramatically. The statistical
analysis of the classical periodogram becomes intractable. The results are
sampling-grid dependent, and no general analysis applicable to all cases has yet
been produced. To simplify the statistical analysis, Scargle proposed that the
definition of the periodogram be modified. His modified periodogram had already been
used by \citet{barning63}, \citet{vanicek69}, and \citet{lomb76}. These authors did
not view the modified periodogram as an attempt to estimate the Fourier power
spectrum from unevenly sampled data, but as a spectral method for searching for the
best-fit harmonic function to their data. The novelty of Scargle's approach was that
he generated the same spectral method as used by these authors by imposing simple
constraints on a generalised form of the Fourier transform. The constraints were
that the modified periodogram should mimic as closely as possible the statistical
properties of the classical periodogram, and that the resulting spectral function
should be insensitive to time translations of the data in the time domain.

The demand that the modified periodogram should mimic as closely as possible the
statistical properties of the classical periodogram was only partially successful.
Forcing time translation invariance, and demanding that the statistics of the random
variable $P_X(\omega)$ at a single selected frequency remain unchanged, that is,
demanding that $P_X(\omega)$ be exponentially distributed, exhausts the free
parameters in Scargle's modified FT, giving Lomb's spectral formula. In this way, he
reproduced some properties of the periodogram for the evenly sampled case. However,
this is the best that he could do. Most other familiar properties are lost. The most
important loss is the existence of independent frequencies.

All relevant information about correlation and mutual dependence of the random
variables $\{ P_X(\omega)\}$ is contained in the window function, $G(\omega)$. (For
a discussion of the window function, see \citet{scargle82}, Appendix D, p 850, and
also his discussion on p 840.) Thus, the coefficient of linear correlation between
$P_X(\omega)$ and $P_X(\omega')$ is given by $G(\omega'-\omega)$ \citep{lomb76}. For
independence of $P_X(\omega)$ and $P_X(\omega')$, it is necessary (but not
sufficient) that $G(\omega'-\omega)=0$. Furthermore, for mutual independence of a
set $\{ P_X(\omega_k): k=1,2,...,r\}$ of periodogram powers, it would also be
necessary (but not sufficient) to have the $\omega_k$ evenly spaced. These are very
difficult conditions to realise in practice. \citet{koen90} has searched numerically
for such mutually uncorrelated sets in a variety of sampling schemes and failed to
turn up more than two simultaneously uncorrelated frequencies.

For Scargle, this loss of independent frequencies is not debilitating. He says (p
840, column 1) that ``... if the frequency grid is well chosen, the degree of
dependence between the powers at the different frequencies is usually small", and (p
840, column 2) that, ``With a wide variety of sampling schemes $G(\omega)$ does have
nulls, or relatively small minima, that are approximately evenly spaced... Such
nulls comprise a set of natural frequencies at which to evaluate the periodogram. At
these frequencies the $P(\omega)$ form a set of approximately independent random
variables - thus closely simulating the situation with evenly spaced data". The
implication, though not explicitly stated by Scargle, is that in spite of the loss
of independence of the random variables $P(\omega)$ at the natural frequencies, the
false alarm probability given by our equation (\ref{scargle-14}) (equation (14) in
\citet{scargle82}, p 839), still provides a reliable significance test in the wide
variety of sampling schemes that he considered.

It seems that Scargle's recommendation for the case of unevenly spaced data is as
follows: evaluate the modified periodogram at the natural frequencies defined by the
given data span, and use the false alarm probability calculated for the evenly
spaced case to evaluate the significance of the periodogram peaks. He further
recommends that, to improve the detection efficiency, we decrease the number of
frequencies inspected (p 842). The effect of this reduction is that we reduce power
threshold for a given significance level of peak-heights.

The value of $N_i$ is a critical ingredient in Scargle's false alarm probability
function. There has been some debate concerning its correct value, as well as its
meaning. \citet{hb86} appear to have been unsatisfied with the value $N_i=[N_0/2]$
and proposed to determine $N_i$ by a method which we describe in the following
section.


\section{HORNE AND BALIUNAS DETERMINATION OF $N_{\lowercase{i}}$}

\citet{hb86}, (HB in the remainder of this paper), determined $N_i$ by the following
procedure. They simulated a large number of data sets, each consisting of
pseudo-Gaussian noise. The periodogram of each data set was evaluated from
$\omega=2\pi/T$ to $\omega=\pi N_0/T$, where $T$ is the total time interval. They
then chose the highest peak in each periodogram, combined these, and fitted the
Scargle false alarm probability function to the peak distribution using $N_i$ as the
variable parameter.

The HB simulations investigated three major types of spacing in the time coordinate.
In the first, the data were evenly spaced in time. In the second, each time followed
the previous one by a random number between 0 and 1. In the third, the data were
clumped in groups of three at each evenly spaced time interval.

In the case where the data are evenly spaced in time, theoretical statistical
analysis provides us with a very clear, unambiguous picture of what to expect from
the simulations: the random variables $\{P_X(\omega_k) : k=1,..., [N_0/2] \}$, where
$\omega_k = 2\pi k /T$ and $T$ is the total time interval covered by the data, are
mutually independent; the window function, which contains all relevant information
about dependencies and correlations of the random variables $P_X(\omega)$, shows
that these are the only frequencies at which the periodogram powers are independent
(Scargle p 840 and 843); the listed frequencies $\omega_k$ contain maximal
information about the power distribution of the sampled signal. This is seen from
the fact that the discrete Fourier transform evaluated at these frequencies contains
exactly enough information to reconstruct completely the original data. So, from
theory, we expect the total number $N_i$ of independent frequencies in the case of
evenly spaced time series consisting of zero mean pure noise to be exactly
$[N_0/2]$. In practice, a simulated time series, generated from a zero mean
distribution, will not have precisely zero mean. We must therefore remove its mean
before finding its periodogram. Once this is done, the theory guarantees that our
simulated data set will have exactly $[N_0/2]$ independent frequencies. The point is
this: for the evenly spaced data sets that we have simulated, the number of
independent frequencies in the periodogram is at most $[N_0/2]$. In real data, this
number may need to be further reduced if we estimate other parameters.

Surprisingly, the best fits obtained by HB consistently produced values of $N_i$
which were substantially higher than this expected upper limit (HB, Table 1, p 759).
In fact, their fitted values are consistently higher than $N_0$, with the exception
of their two smallest data sets (10 and 15 points respectively) where the fitted
value of $N_i$ is slightly less than $N_0$, but still about twice as large as
expected.

These results are puzzling. Theory and simulations appear to be in conflict.
\citet{cumming99} note that Baliunas has indicated typographical errors in the
values listed in HB. \citet{koen90} and \citet{sc96} have also noted mistakes in HB.
We have repeated the HB simulations for the case of even sampling in the time
domain. We have also extended somewhat the scope of their investigations to consider
the alternative false alarm probability function proposed by \citet{sc98}, as well
as the effects of over-sampling the periodogram. The results are interesting, and we
report them in the corresponding sections below.

\begin{figure}
\includegraphics*[width=84mm,height=55mm]{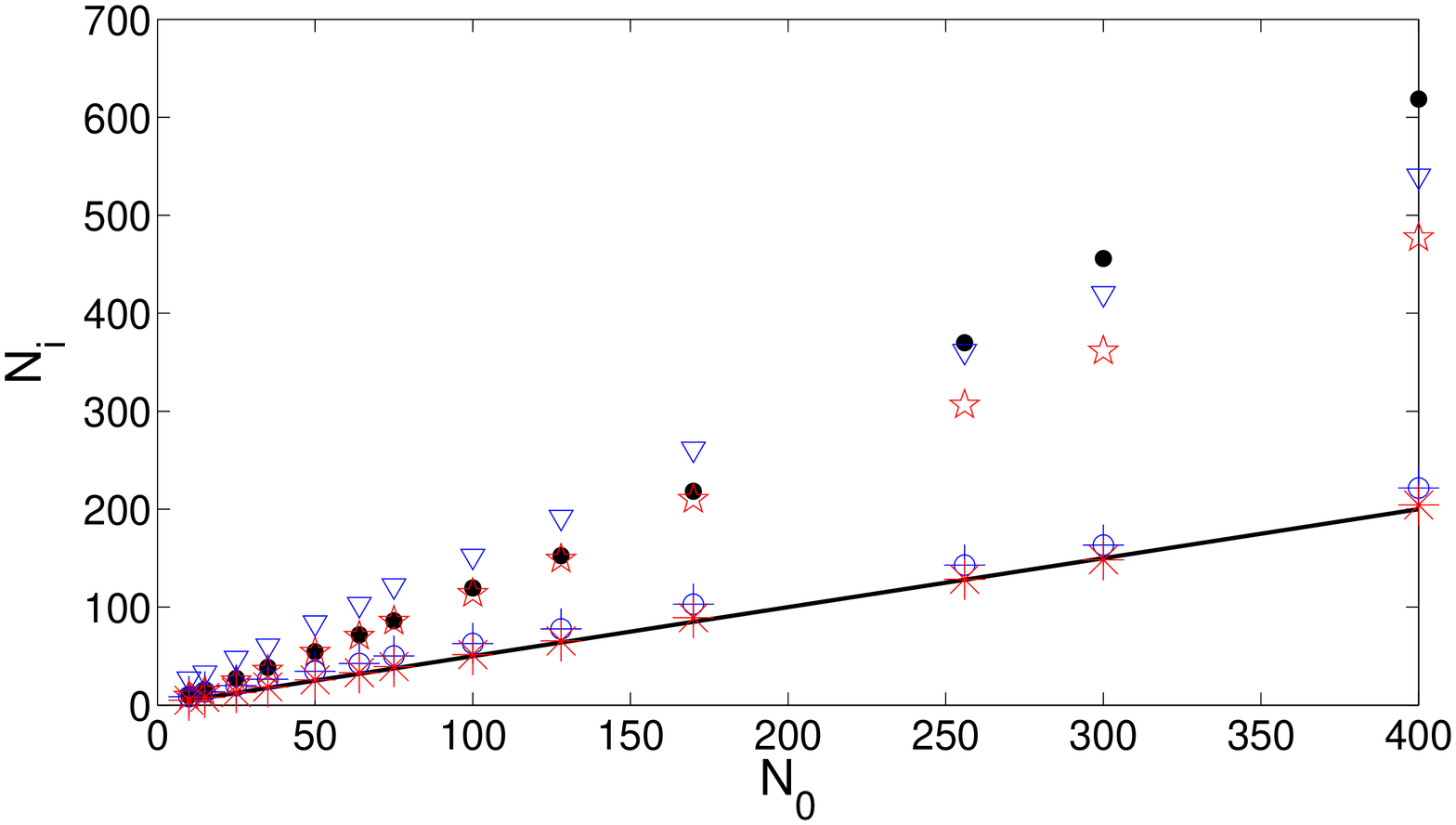}\\
 \caption{
Plots of $N_i$ vs. $N_0$ of the data published by \citet{hb86} in their Table 1, p
759, for the case of evenly spaced data, and of our simulations, fitting the Scargle
and Schwarzenberg-Czerny false alarm functions to the empirical CDF's obtained by
sampling at the natural frequencies and by over-sampling. Solid dots: published
Horne and Baliunas values; asterisks: Scargle function fitted to empirical CDF's
obtained by sampling at the natural frequencies; circled crosses:
Schwarzenberg-Czerny function fitted to the same; stars: Scargle function fitted to
empirical CDF's obtained by over-sampling; triangles: Schwarzenberg-Czerny function
fitted to the same. The solid line is the theoretically expected relationship
$N_i=[N_0/2]$.
   }
 \label{figure0}
\end{figure}

In our first set of simulations, we attempted to reproduce the results reported by
HB in their Table 1, p 759, for the case of evenly spaced data. HB describe the
method they followed in their simulations as follows: ``The periodogram of each data
set was evaluated from $\omega=2\pi/T$ to $\omega=\pi N_0/T$ ... The highest peak
was then chosen in each periodogram." It was not clear to us whether they sampled
the periodogram values $P_X(\omega)$ {\em only} at the natural frequencies $\omega_k
= 2\pi k /T$, and then chose the highest periodogram power from this restricted
sampled set, as prescribed by Scargle; or whether they followed the practice of a
not insubstantial number of astronomers who search for the highest periodogram peak
in the given range by grossly over-sampling the periodogram, and then choose the
maximum value obtained irrespective of whether it occurs at one of the natural
frequencies $\omega_k$. Accordingly, we ran two sets of simulations implementing
both procedures. We fitted the Scargle false alarm function to our results by the
method of least squares. All our periodograms were normalised using the sample
variance of the simulated data, and not the variance of the distribution used to
generate the sample. We failed to reproduce the HB results in detail. Sampling the
periodogram at the natural frequencies only and choosing the highest value among
these yielded values of $N_i$ that were consistently lower than those obtained by
HB. In fact, we obtained values very close to $[N_0/2]$, as expected theoretically,
but in conflict with the results published by Horne and Baliunas. Searching for the
highest peak by over-sampling also yielded values that were consistently lower than
HB, but higher than sampling at the natural frequencies. More precisely, our results
agree closely with those of HB for the smaller data sets up to $170$ data points.
This leads us to suspect that the HB table was constructed by gross over-sampling.
However, our results strongly deviate from theirs for the larger data sets with $N_0
> 170$, with our values being substantially lower. Plotting $N_i$ vs. $N_0$ (Figure
\ref{figure0}),
 we observe the following features. The values yielded by our simulations increase
linearly with $N_0$, as expected. In contrast, the results published by HB in their
Table 1 appear to lie, not on a quadratic (as claimed by them), but on two straight
lines of different slope, a sharp change in slope appearing for data sets with $N_0
> 170$. This seems to be indicative of a systematic error. Fitting a quadratic function to
these data points, as was done by HB, may therefore be misleading and renders
suspect its use in estimating the parameter $N_i$.

Note however that, in the case of over-sampling, both our results and those of HB
consistently yield values of $N_i$ that are higher than the theoretically expected
value of $[N_0/2]$. These values are thus apparently in conflict with the theory.
The interpretation of $N_i$ as the number of independent frequencies is therefore
questionable in this context. The HB method for determining $N_i$ is eminently
practical and reasonable, but {\em it only yields correct values when the
periodogram is sampled at the natural frequencies}. This means that, in the context
of over-sampling, we cannot assign to the parameter $N_i$ the meaning that it had in
its original derivation, namely the number of independent frequencies in the
associated periodogram. Rather, we must treat $N_i$ as nothing more than a floating
parameter in a one-parameter family of candidate CDF functions which we are
attempting to fit to our data.

\begin{figure}
\includegraphics*[width=80mm,height=48mm]{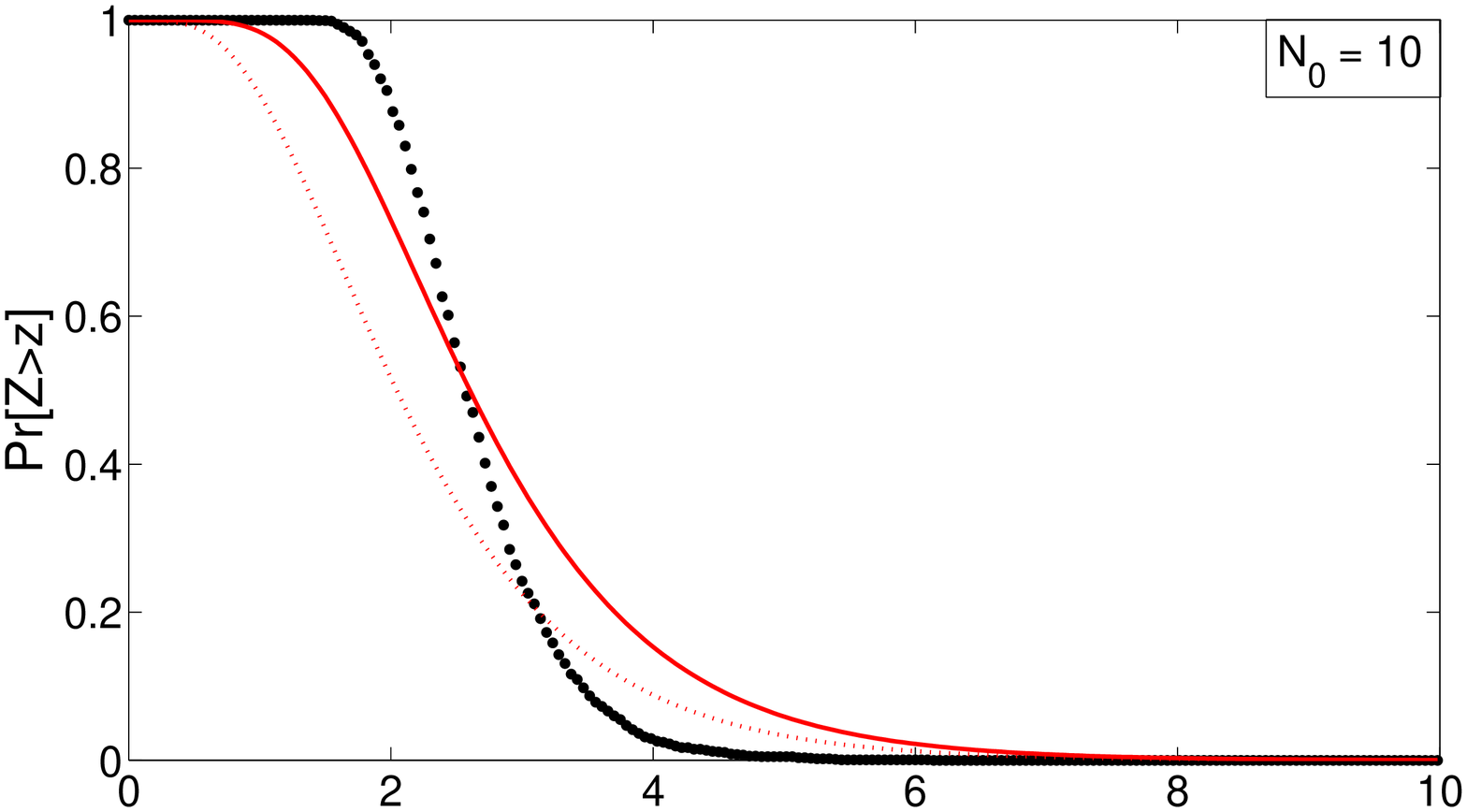}\\
\includegraphics*[width=80mm,height=48mm]{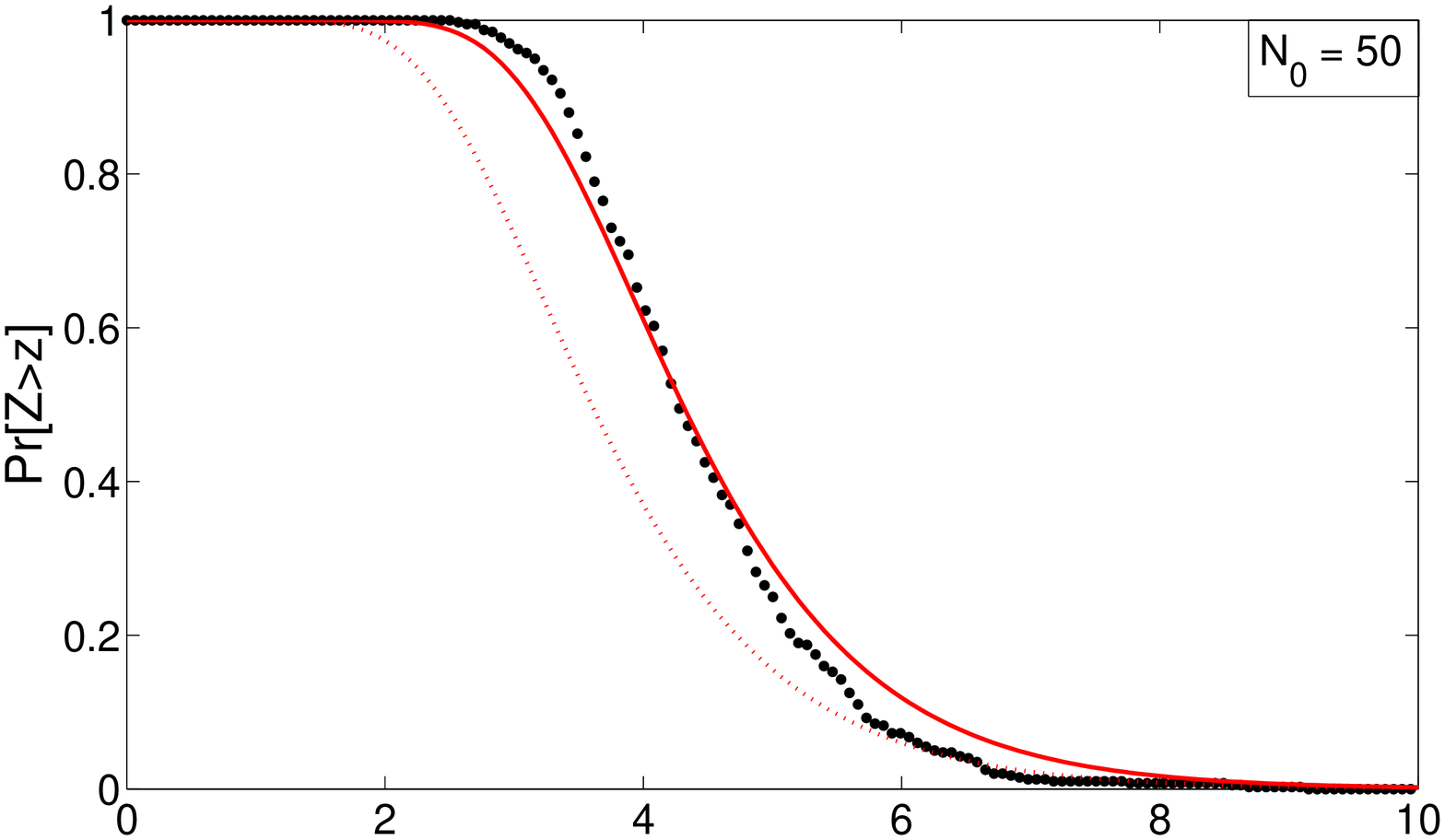}\\
\includegraphics*[width=80mm,height=48mm]{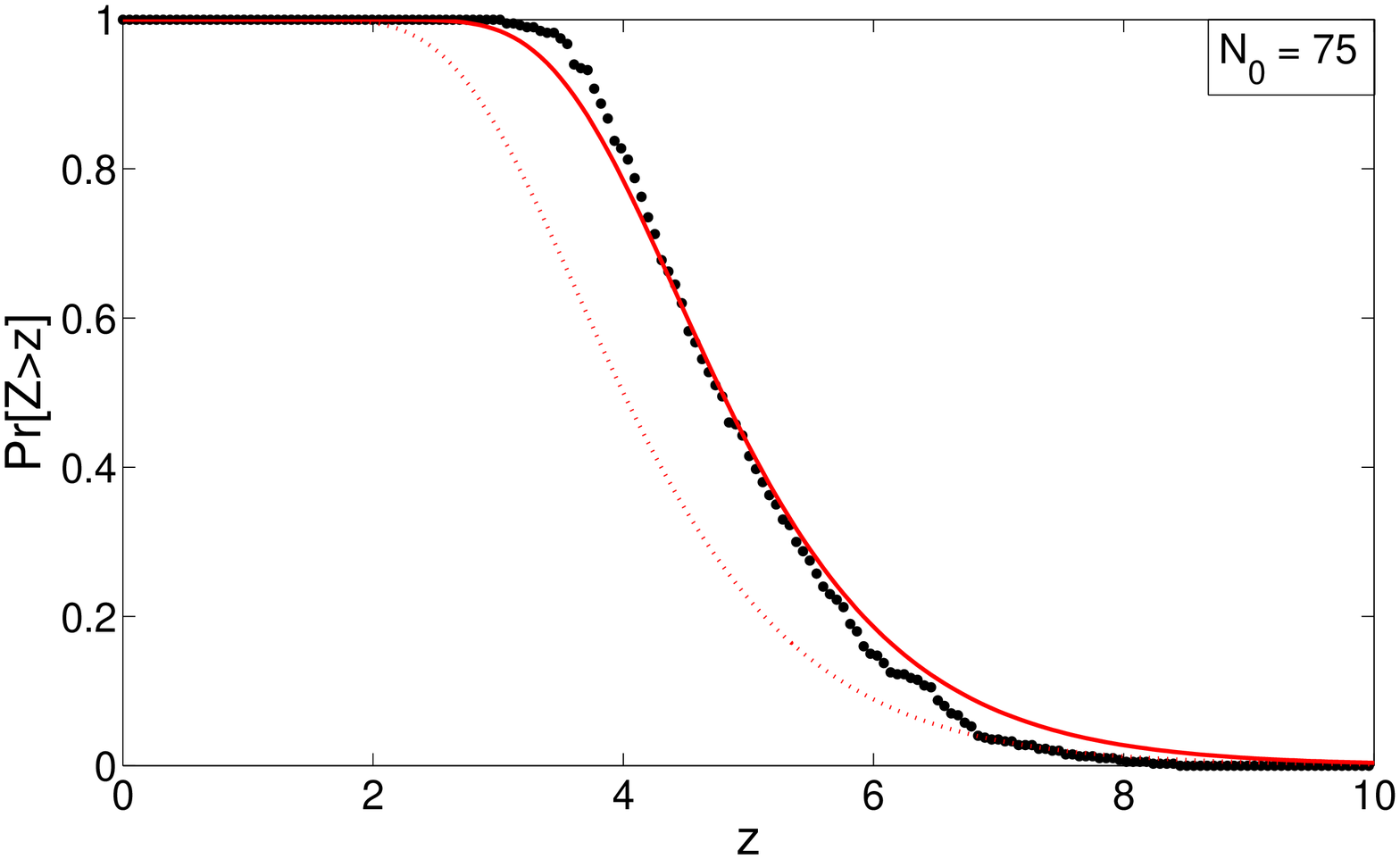}\\
 \caption{
Empirical CDF's (heavy dotted line) constructed by over-sampling the periodogram,
with best-fitting Scargle false alarm probability function (solid line) for (a)
$N_0=10$, (b) $N_0=50$, and (c) $N_0=75$ data points. The fit improves with
increasing $N_0$. The corresponding best-fit values of $N_i$ are (a) $N_i=9.09$, (b)
$N_i=54.00$, and (c) $N_i=85.82$. The light dashed line shows the Scargle function
for $N_i=[N_0/2]$. In all cases the best-fit value of $N_i$ exceeds $[N_0/2]$.
   }
 \label{figure1}
\end{figure}

Another problem with the HB method should be noted. Inspection of a plot of the
best-fit Scargle false alarm probability function shows it to be a very
uncomfortable fit to the experimentally obtained cumulative distributions of
periodogram peak heights (see Figure \ref{figure1}). This is true both in the case
of sampling at the natural frequencies and of over-sampling. Its general trend is
good: it is flat near value 1 at low peak heights, drops rapidly over the peak
height mid-range, and levels off to zero for larger peak heights. However, its
detailed behaviour simply does not match that of the experimental curve. It drops
too quickly, and levels off too soon. This mismatch is most pronounced for small
data sets, and becomes progressively less noticeable as the data sets increase in
size. But it never vanishes completely. The conclusion forced on us by our
simulations is that {\em the Scargle false alarm probability function fails to
reproduce the detailed behaviour of the simulated data sets}. This is both good news
and bad news: good news because it shows that the Scargle function {\em
underestimates} the significance of periodogram peaks; and bad news because it
leaves us without an useable false alarm probability function.

In summary, we cannot in general regard $N_i$ as anything more than a fitting
parameter. Furthermore, {\em the Scargle probability function incorrectly describes
the statistical behaviour of the periodogram in these simulations}. We discuss a
possible reason for its failure in a later section. For the moment, we simply note
that it manifestly fails to provide a convincing fit to the empirical CDF produced
by our simulations. It displays the correct general characteristics of a CDF but,
notoriously, all CDF's tend to look alike, so simply displaying correct general
features is not a point in its favour. Our conclusion therefore is that {\em the HB
method is not in general a way to assess the number of independent frequencies in a
periodogram. Rather, it is a method for estimating the best-fit parameter $N_i$ in
an ill-fitting class of candidate CDF functions}.

This does {\em not} make the Scargle probability function or the HB method for
estimating $N_i$ worthless. In those cases (large data sets) where the Scargle
function gives a reasonable fit to the empirical data, the HB method provides a
value of $N_i$ that makes the Scargle function a good estimate of the correct false
alarm probability and provides a formula in closed form that can be used as a
significance test. However, note that this formula consistently underestimates the
significance of periodogram peaks, this underestimation becoming increasingly severe
as the data sets become smaller.

\begin{table*}
\small
 \centering
 \label{TabulatedValues}
 \begin{minipage}{140mm}
\caption{Results of Monte Carlo simulations} \vspace*{4mm}
  \begin{tabular}{@{}ccccccc@{}}
    \hline
 &&&\multicolumn{2}{c}{\bf Over-sampled}&\multicolumn{2}{c}{\bf Natural Frequencies} \\
 $N_0$ & HB Value&Number   &Scargle Function  & SC Function &Scargle Function
 & SC Function  \\
        &of $N_i$&of Tests & Best-fit $N_i$  &Best Fit $N_i$& Best Fit $N_i$
        & Best Fit $N_i$ \\
 \hline
 10.00&9.70&1395.00&9.09&26.80&5.00&8.70 \\
 15.00&14.45&347.00&14.09&32.80&7.90&13.60 \\
 25.00&27.38&213.00&24.91&47.80&12.80&20.00  \\
 35.00&38.40&214.00&35.64&60.40&18.50&26.50  \\
 50.00&54.45&369.00&54.00&84.00&25.70&34.40  \\
 64.00&71.76&512.00&70.45&102.80&33.00&42.40  \\
 75.00&86.05&153.00&85.82&121.90&39.30&50.10  \\
 100.00&119.58&296.00&113.91&152.10&51.60&62.80  \\
 128.00&152.53&913.00&149.09&191.80&65.50&77.70  \\
 170.00&218.33&218.00&210.09&261.40&89.30&103.00  \\
 256.00&369.97&224.00&306.36&361.20&128.40&143.00 \\
 300.00&455.95&107.00&361.45&420.20&148.40&163.30  \\
 400.00&618.69&106.00&477.18&540.10&204.30&221.60 \\
 \hline
\end{tabular}
\tablecomments{ Comparison of Horne and Baliunas values of $N_i$ with the results of
our numerical simulations, fitting both Scargle and Schwarzenberg-Czerny (SC) false
alarm functions to CDF's constructed from over-sampled periodograms and from
periodograms sampled at the natural frequencies. The corresponding best-fit
functions are displayed in Figures \ref{figure1} and \ref{figure2}, together with
the corresponding functions constructed with the correct value of $N_i=[N_0/2]$.}
\end{minipage}
\end{table*}

\normalsize
\section{THE SCHWARZENBERG-CZERNY FALSE ALARM FUNCTION}

\citet{koen90} pointed out an important implicit assumption in Scargle's derivation
of his false alarm probability function. Scargle assumed that the variance
$\sigma_X^2$ of the data $X_k$ is known {\em a priori}. There are situations in
which this condition is true, but it is satisfied neither in the case of real
astronomical data nor in that of the HB simulations. In the simulations, pseudo-data
are generated using a preselected variance and mean (chosen to be zero), but the
variance and mean of the generated sample will differ in general from those used in
their generation. Thus both variance and mean need to be estimated from the data.

This changes the statistical analysis significantly. \citet{sc98}, in a particularly
clear and thorough exposition of the issues involved, has shown that the CDF of
maximum peak heights appropriate to the Lomb-Scargle periodogram and calculated from
a finite sample of Gaussian pure noise, is the (regularised) incomplete beta
function
\begin{equation}
I_{1-z/[N_0/2]}([N_0/2],1)=
 \left ( 1-\frac{ z }{[N_0/2]} \right )^{[N_0/2]}
    \label{SC-1}
\end{equation}
To construct the corresponding false alarm probability function, we need to use this
distribution in place of the exponential distribution used above. If in our
periodogram we can identify a set of frequencies at which the periodogram powers are
mutually independent, then the probability that the power at at least one of these
frequencies rises above given threshold power $z$ is given by
\begin{equation}
 {\rm Pr\ }[Z_{\rm max} > z ]
 = 1- \left[ 1 - \left(1-\frac {z/\sigma^2_X} {[N_0/2]} \right )^{[N_0/2]} \right]^{N_i}
     \label{SC-2}
\end{equation}
where $N_i$ is the number of mutually independent frequencies inspected, and $Z_{\rm
max} = {\rm sup\ }\{ Z_1, Z_2, ... , Z_{N_i} \}$ is the maximum power among the
mutually independent powers $Z_\mu$. In our discussion, we shall call equation
(\ref{SC-2}) the {\em Schwarzenberg-Czerny false alarm probability function}. In
passing, note that \citet{sc98} provides a number of alternative distributions and
test statistics appropriate to other methods of data analysis and is able thereby to
resolve extant disputes about the ``correct" normalisation procedure for
periodograms.

In the limit $N_0 \rightarrow \infty$, the distribution in equation (\ref{SC-1})
becomes exponential and coincides with that used by Scargle. Accordingly, in the
same limit, the associated false alarm probability function in equation (\ref{SC-2})
reduces to the Scargle false alarm function. A $Q-Q$ plot of the
Schwarzenberg-Czerny vs. Scargle false alarm functions (see \citet{sc98}, Figure 1,
p 835) shows that, while the agreement between them is good for large $N_0$, they
differ substantially for small data sets, with Schwarzenberg-Czerny's false alarm
function yielding consistently smaller false alarm probabilities than Scargle's.
According to this analysis, therefore, for given $N_i$, the Scargle false alarm
function consistently {\em underestimates} the statistical significance of
periodogram peaks.

\begin{figure}
\includegraphics*[width=80mm,height=48mm]{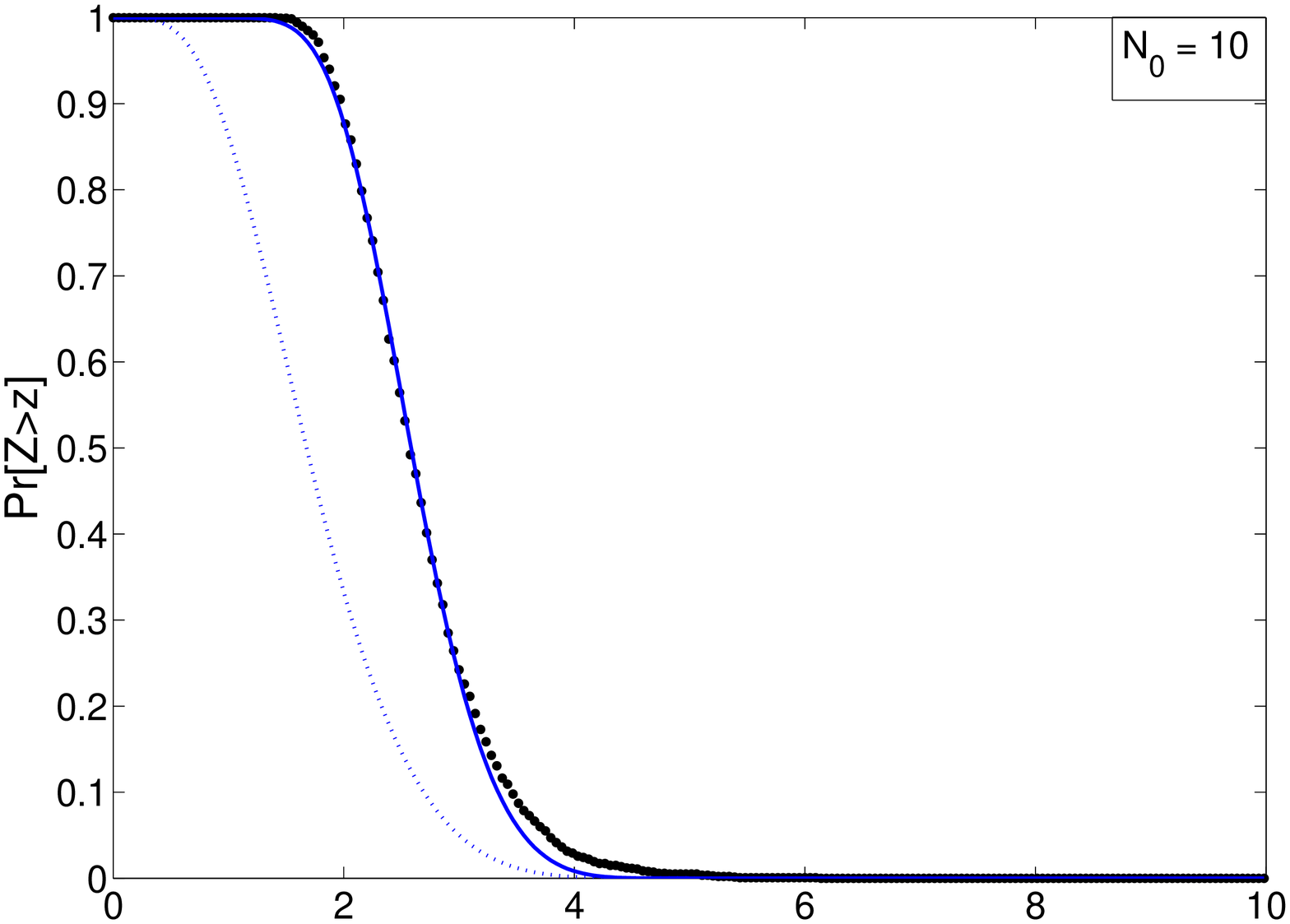}\\
\includegraphics*[width=80mm,height=48mm]{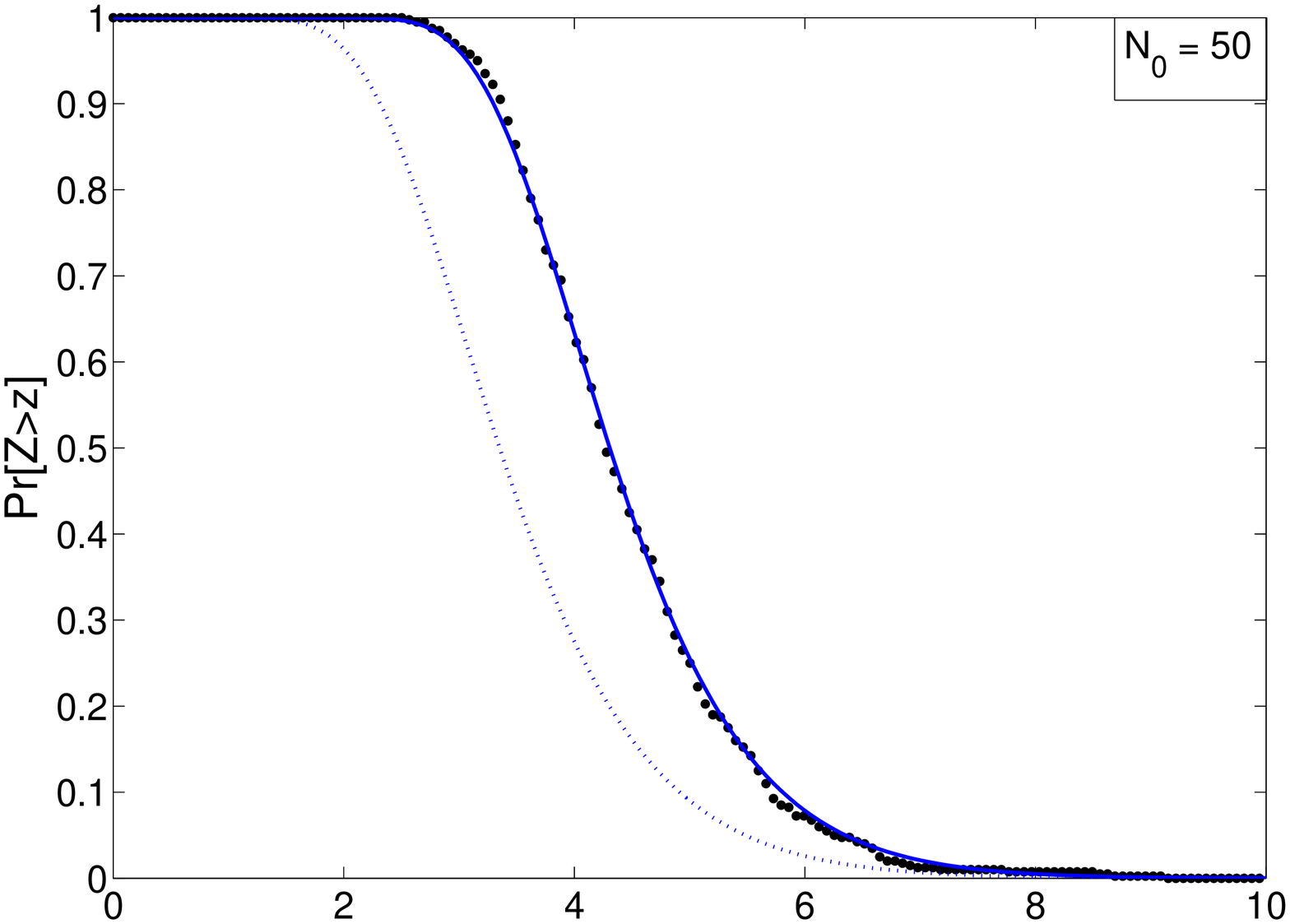}\\
\includegraphics*[width=80mm,height=48mm]{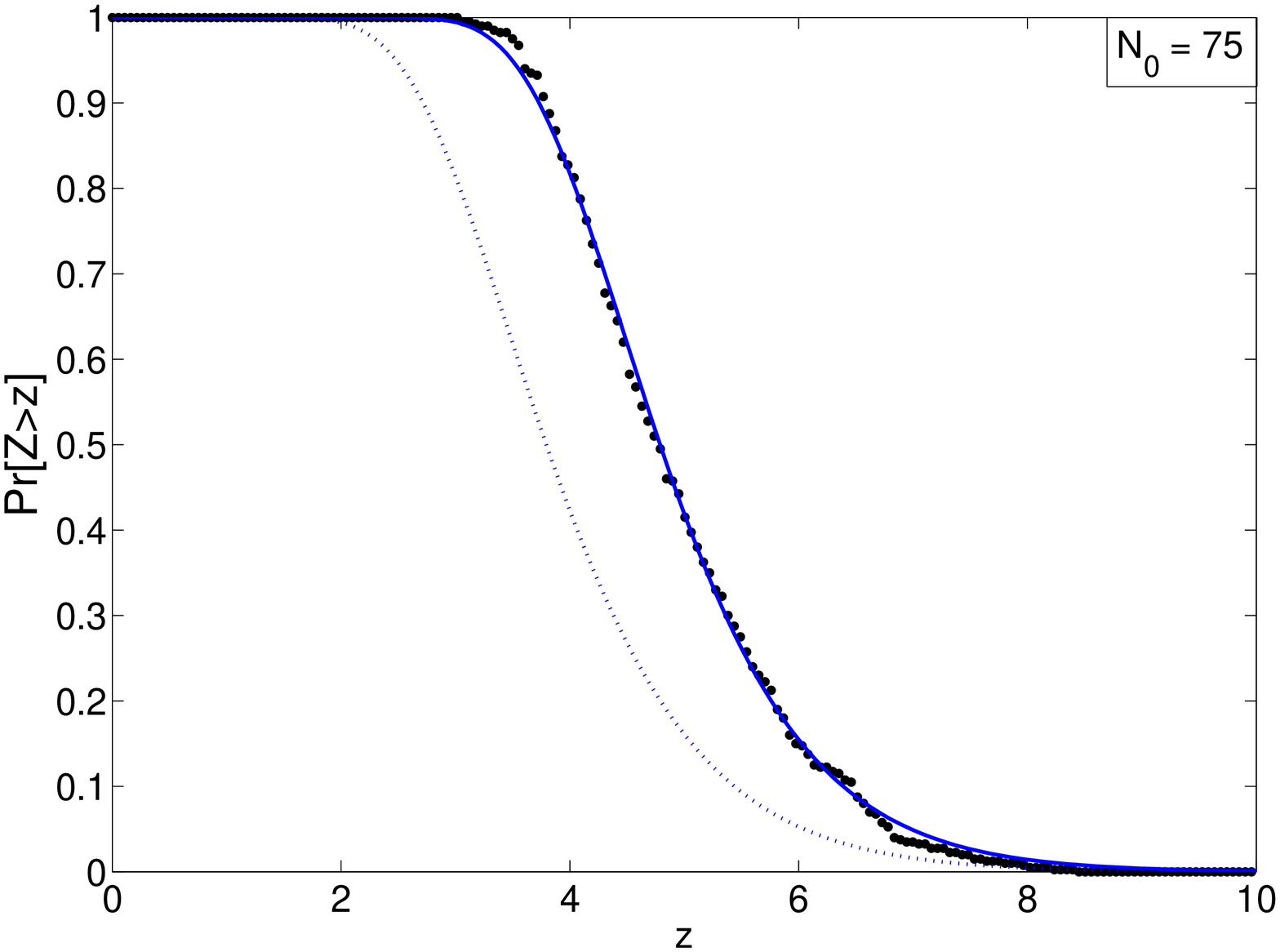}\\
\caption{Empirical CDF's (heavy dotted line) constructed by over-sampling the
periodogram, with best-fitting Schwarzenberg-Czerny false alarm probability function
for (a) $N_0=10$, (b) $N_0=50$, and (c) $N_0=75$ data points.
 The fits are significantly better than the corresponding ones for Scargle's function.
 However, for low $N_0$, Schwarzenberg-Czerny's distribution is still significantly
 different from the empirical one and overestimates the significance of high peaks.
The light dashed line shows the corresponding Schwarzenberg-Czerny false alarm
function for $N_i=[N_0/2]$. In all cases, the best-fit value of $N_i$ again exceeds
$[N_0/2]$.}
 \label{figure2}
\end{figure}

One reason for the failure of the Scargle function to reproduce the behaviour of our
empirical CDF's may be its implicit assumption that the variance $\sigma_X^2$ is
known {\em a priori}. To correct this error, we replaced the Scargle function by
Schwarzenberg-Czerny's and repeated the HB simulations for equally spaced data.
Using their method for determining $N_i$, we fitted the Schwarzenberg-Czerny false
alarm function to our empirical CDF's. We found very good, but not perfect,
agreement between the best-fit theoretical curves and the corresponding empirical
ones, with the greatest deviations occurring for small data sets (See Figure
\ref{figure2}).
For these, the theoretical best-fit curves consistently yield values that are {\em
lower} than those of the empirical curves, thus {\em overestimating} the
significance of peaks. For the larger values of $N_0$, the deviations of the fitted
from the empirical curves may be understood in the context of order statistics.

In spite of the excellent nature of these fits, there is nevertheless an interesting
feature in these results that is worth noting. For the CDF's of periodogram powers
sampled at the natural frequencies, {\em the best-fit values of $N_i$ are
consistently larger than the theoretically expected number of independent
frequencies, which is at most $[N_0/2]$} (see Table 1). Correspondingly, a plot of
the Schwarzenberg-Czerny function for the value $[N_0/2]$ of independent frequencies
yields a curve that deviates badly from the corresponding empirical CDF and which
leads to a severe {\em overestimation} of the significance of periodogram peaks. We
have no option but to conclude from these results that, like the Scargle false alarm
function, the Schwarzenberg-Czerny false alarm function, given by equation
(\ref{SC-2}), appears not to describe the CDF's of our simulations. Note also from
Table 1 that the best-fit values of $N_i$ for CDF's constructed from over-sampled
periodograms are higher than those for the CDF's obtained by sampling at the natural
frequencies. This is consistent with our previous results for the Scargle false
alarm function. The results of our simulations again appear to be at variance with
the theory. For evenly spaced data, the theory (which seems unassailable) predicts
unambiguously the existence of at most $[N_0/2]$ independent frequencies, with the
CDF for periodogram powers sampled at these frequencies given by the
Schwarzenberg-Czerny false alarm function. Our empirical CDF's differ substantially
from those predicted by this theory, with HB best-fits occurring at values of $N_i$
that are consistently higher than expected. These results force us to the following
conclusions. First, when using the HB method to estimate $N_i$, {\em we cannot
interpret the best-fit value of $N_i$ as the number of independent frequencies in
our periodogram. Rather, we must treat $N_i$ as a floating parameter in a
one-parameter family of candidate CDF functions}. This conclusion is consistent with
that stated in the previous section. Second, as candidate CDF functions, the
Schwarzenberg-Czerny false alarm function appears to be superior to Scargle's.

\section{FALSE ALARM FUNCTIONS FOR UNEVENLY SPACED DATA}

The principal difficulty encountered when searching for a false alarm function in
the case of unevenly spaced data is the loss of the so-called independent
frequencies. This loss is not apparent. It is real. The problem is not that they are
difficult to identify but nevertheless present. It is that they are not there at
all, except perhaps for a small set that can be counted on the fingers of one hand.
A significance test based on so small a number of independent frequencies is not
useful. It would require us to sample the periodogram at no more than a few
frequencies, making it highly likely that we would miss most of the significant
periodicities in our data because of sparse sampling.

The loss of a sizeable set of mutually independent frequencies puts us into a
difficult, possibly intractable, position vis-a-vis the search for a theoretical
formula in closed form for false alarm probabilities. Were such a formula available,
it would certainly be a great boon. Realistically however, it seems unlikely that
such a formula could ever be found for the general case of arbitrarily spaced data.

The only alternative to a theoretical false alarm function is an empirically
generated one. \citet{sc98} expresses a distinct lack of confidence in this
approach. He states, p 832, ``We consider the opinion that all statistical problems
related to the periodograms can be solved by Monte Carlo simulations to be
over-optimistic". His skepticism regarding simulations is due to the unreliability
of random number generators. He says, p 832, ``The simulations have problems of
their own, related chiefly to the untested effects of the discrete random number
generators and periodogram algorithms on the tails of the continuous distributions",
and again on p 833, ``The Monte Carlo simulations rely on rare events of low
probability, for which neither the accuracy of random number generators nor the
accuracy of periodogram algorithms is well tested".

The strong sentiments expressed by Schwarzenberg-Czerny offer little cheer to
observers, whose principal need is a reliable method for assessing candidate
periodogram peaks. The current generation of theoretical distributions are all based
on the assumption of independent frequencies and all require a value of $N_i$. In
the case of evenly spaced data, it might be argued that the correct value for $N_i$
is $[N_0/2]$, suitably reduced by the number of parameters already estimated from
the data. For the general case however, even were we to believe the conjecture that
independent frequencies exist, there appears to be no clear {\em a priori}
theoretical criterion for choosing the value of $N_i$, and the only practical method
offered is that of HB in which we fit some chosen theoretical distribution to the
simulated CDF's. Necessity therefore forces us, against Schwarzenberg-Czerny's
advice, into the route of Monte Carlo simulations.

The realisation that the conjecture of the existence of independent frequencies is
false forces us to re-evaluate both the role of Monte Carlo simulations and the use
of theoretical false alarm probability functions. Schwarzenberg-Czerny's opinion
regarding random number generators is not unwarranted. However, the performance of
random number generators is continually being improved. There is every reason to
believe therefore that existing problems with random number generators will
eventually be resolved. In contrast, the problem of the lack of independent
frequencies is permanent. The Monte Carlo simulation option is therefore not as
bleak as may first appear. As regards the use of theoretical false alarm probability
functions, we do not really need them. The empirically generated CDF's contain all
the information that we need, whether or not we have a closed-form formula for them,
and can be used to determine significance thresholds. A closed-form formula would be
useful insofar as it facilitates calculation of the thresholds, but is not
essential. If one is needed, we can resort to fitting the empirical CDF as closely
as possible by {\em any} suitable form of trial function. In fact, we do not need
even to fit the entire CDF. We are interested only in the high-peak tail above a
certain minimum confidence threshold and so need only obtain a good fit in that
region. Should formulae be needed for other regions, we can resort to multiple fits
that together cover the entire CDF.


\section{THE PROBLEM OF OVER-SAMPLING}

Theoretical false alarm probability functions are based on the assumption of the
existence of independent frequencies and contain the number $N$ of frequencies
inspected as a parameter. When the periodogram is inspected at the maximum number
$N_i$ of independent frequencies, $N=N_i$. For example, Scargle's function is given
by
\begin{eqnarray}
  {\rm Pr} [ Z > z ] = 1 - F_{Z_{\rm max}} (z) = 1 - \left( 1- e^{-z} \right)^{N}
\end{eqnarray}
when $N$ is the number of independent frequencies {\em inspected}.

\begin{figure}
\includegraphics*[width=84mm,height=55mm]{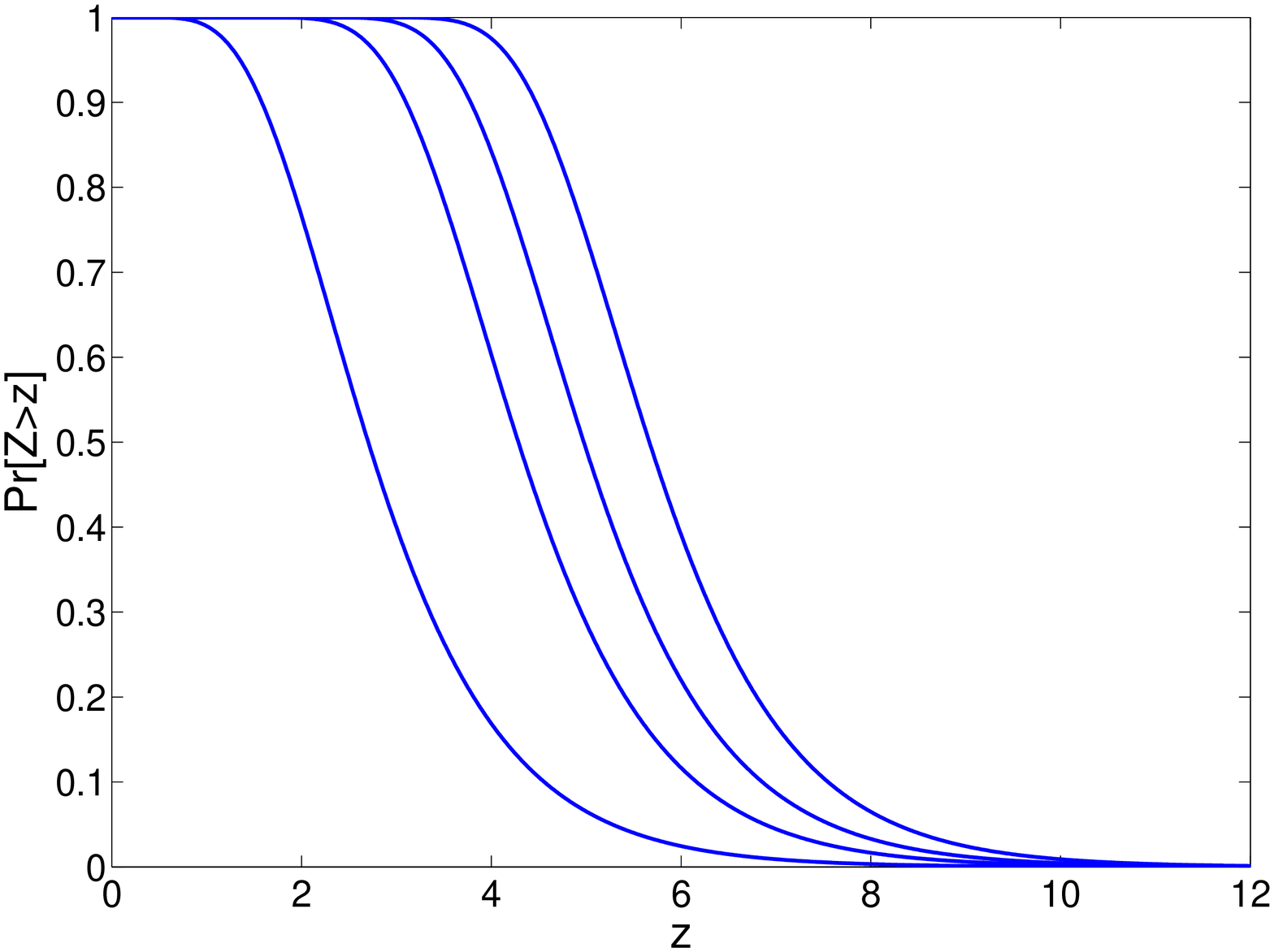}\\
 \caption{
Scargle false alarm probability function as a function of $N$ for values $N= 10, 50,
100, 200$. As $N$ increases, the probability of finding a peak above any given
threshold value increases. This illustrates Scargle's `statistical penalty': if many
independent frequencies are inspected for a spectral peak, we should expect to find
a large peak even when no signal is present. As $N$ increases, the CDF moves
progressively to larger peak-height values without limit.}
 \label{figure3}
\end{figure}

From Figure \ref{figure3}, it is seen that, for given $z$, this probability
increases as the number $N$ of sampled independent frequencies is increased.
Scargle, p 839, describes this property as the {\em statistical penalty} that we
must pay for inspecting a large number of frequencies. He explains this by saying
that ``if $N$ independent experiments are carried out, even if each one has a very
small probability of succeeding, the chance of one of them succeeding is very large
if $N$ is large enough (approaching certainty as $N$ approaches infinity)." He also
notes that the expected value of the maximum power $Z_{\rm max}$ of a white noise
spectrum over a set of $N$ frequencies at which the power is independent is given by
\begin{eqnarray}
  \langle Z_{\rm max} \rangle = \sum_{k=1}^N \ \frac{1}{k}
\end{eqnarray}
which diverges logarithmically with $N$.

These comments appear alarming. At face value, they seem to suggest that prodigious
sampling of the periodogram at the independent frequencies might lead eventually to
the dismissal of {\em all} periodogram peaks as spurious. They also appear strongly
to discourage over-sampling of the periodogram in an attempt to pin down more
precisely the frequency of a periodicity. Indeed, their effect has been so strong on
some that they refuse to evaluate periodogram power at any frequencies other than a
selected subset of the ``natural frequencies". Were these extreme conclusions drawn
from Scargle's comments correct, the periodogram method for searching for
periodicities would be severely compromised.

To understand Scargle's comments correctly, we need first to note that the false
alarm function is deduced assuming that we are able to identify $N$ independent
frequencies $\omega_k$. An evenly sampled time series consisting of $N_0$ data
points guarantees the existence of {\em at most} $N=[N_0/2]$ mutually independent
frequencies, namely the ``natural" ones. There can be no more. The original data set
can be fully recovered from the DFT at these frequencies. So the information
contained in periodogram powers at all other frequencies cannot be independent of
these. For a given evenly sampled time series consisting of $N_0$ points, there is a
maximum value of $N$ at which we can sample the periodogram independently, namely
$[N_0/2]$, and hence a maximum value of $\langle Z_{\rm max} \rangle$. In practice
therefore, there is no logarithmic divergence to fear.

\begin{figure}
\includegraphics*[width=84mm,height=55mm]{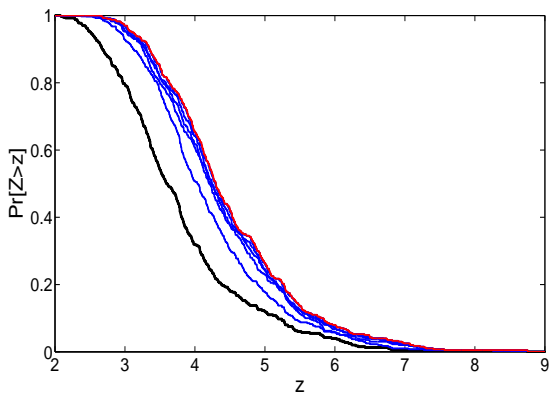}\\
 \caption{Empirical CDF's as a function of over-sampling. The figure shows the CDF's
 corresponding to sampling at 1, 2, 3, 4, 5, and 10 times the Scargle sampling rate.
 The corresponding CDF's converge rapidly to a limiting CDF. The limiting CDF
 coincides almost perfectly with the CDF for an over-sampling factor of 10.}
 \label{figure4}
\end{figure}

Second, if we over-sample the periodogram, the CDF is no longer correct. The powers
at the sampled frequencies are no longer independent, and so equation
(\ref{scargle-14}) ceases to be the correct description of the distribution. In
these circumstances, it is not useful to look for a theoretical formula for the CDF.
Even if it were mathematically tractable, it probably would not be worth the effort
of obtaining an expression in closed form for it. The difficulties in obtaining a
formula for this CDF, however, do not prevent us from obtaining an excellent
approximation to it through numerical experiments. The results of our simulations in
this respect are encouraging. Successive over-samplings produce progressively less
effect on the CDF, until {\em it eventually converges to a limiting CDF beyond which
no further refinement of the sampling grid changes the result.} (See Figure
\ref{figure4}.)
With hindsight, we should have expected this. The original time domain data contain
a finite amount of information. There is therefore a limit to how much information
they can be forced to yield.

Based on the numerical experiments described in this paper, we would therefore like
to refine Scargle's lesson, drawn from a consideration of statistical penalties. The
(gloomy) lesson he drew was: ``If many frequencies are inspected for a spectral
peak, expect to find a large peak power even if no signal is present" \citep[p 840,
column 1]{scargle82}. Our revision of Scargle's lesson is this: {\em If many
frequencies are inspected for a spectral peak, expect to find a large peak power
even if no signal is present - but the total number of independent frequencies
present in any given time series is limited, so don't expect the number of large
peaks produced by white noise to increase without limit. More importantly, {\bf \em
over-sampling the periodogram does not dramatically increase the number of large
peaks expected}}.


\section{A PRACTICAL METHOD FOR DETERMINING FALSE ALARM PROBABILITIES}

The theoretical false alarm probability functions extant in the literature all rely
for their validity on the existence of independent frequencies. Such a set is
guaranteed for evenly spaced data, but not for data that are unevenly spaced. Even
in the case where the data are evenly spaced, we may wish to inspect the periodogram
at frequencies that do not coincide with Scargle's natural ones. Such is the case
when a pronounced peak occurs at an intermediate frequency. How do we assess the
significance of periodogram peaks in these cases?

Based on our investigations described above, we suggest the following method:
 \begin{enumerate}
 \item
{\em Using the sampling times of the actual data set to be analysed}, construct a
large number of pseudo-Gaussian random time series.

\item
Select a convenient grid of frequencies that cover the frequency range in the
periodogram that is to be inspected. (We discuss how to choose these frequencies in
the next paragraph. For the moment, assume that they have been selected.)

\item
Construct the periodogram for each pseudo-random time series, sampling it at
each of the selected frequencies.

\item
In each periodogram, identify the highest periodogram power that occurs {\em at the
pre-selected frequencies only}, and use these highest values to construct the CDF of
these highest power values.
 \end{enumerate}
The CDF thus obtained is an empirically generated graphical representation of the
probability function ${\rm Pr}[Z_{\rm max}\leq z]$. It gives the probability that
pure noise alone could have produced power values less than or equal to a given
threshold
value $z$ at each of the selected sampling frequencies.\\

\begin{minipage}[h]{7.4cm}
{\normalsize \bf The plot of $1-{\rm Pr}[Z_{\rm max}\leq z]$ is thus the required
false alarm probability function. It gives the probability that pure noise alone
could produce a peak {\em at the inspected frequencies} of value higher than the
threshold $z$. \\}
 \end{minipage}

\begin{figure}
\includegraphics*[width=84mm,height=48mm]{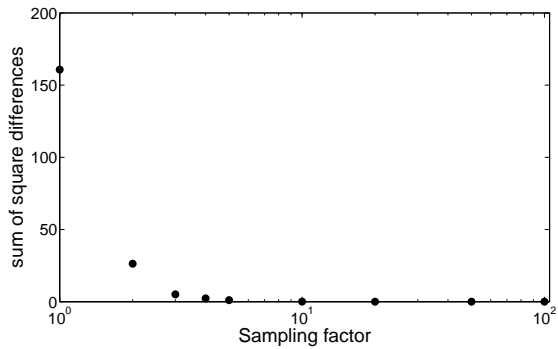}\\
 \caption{
Logarithmic plot of the sum of square deviations (from the limiting CDF) of the CDF
for $\nu$ times over-sampling vs. the over-sampling factor $\nu$. The convergence to
the limiting curve is seen to be very rapid. For the data set used in this
simulation, the convergence occurs approximately at an over-sampling factor of 5.
The convergence shows up in this plot as a sharp levelling off of the graph.}
 \label{figure5}
\end{figure}

How do we choose the frequencies at which to sample the periodogram? In a sense, it
makes little difference how we choose them since, once chosen, we generate an
empirical false alarm probability function that is tailor-made for our particular
choice. However, for each choice, there is a price to be paid, and the final
decision on how to choose the sampling frequencies is determined by what we consider
to be the best compromise between the price paid and the advantage gained. For a
given false alarm probability $p_A$, the denser the sampling, the higher the
associated threshold $z$, with the heaviest penalty being paid for over-sampling
sufficiently dense as to produce a fully resolved periodogram curve. In our
simulations, this occurred at approximately five times the Scargle sampling rate,
that is, using $\Delta \omega = \frac{1}{5}(\pi N_0/T)$. (See Figure \ref{figure5}.)
\begin{figure}
\includegraphics*[width=84mm,height=48mm]{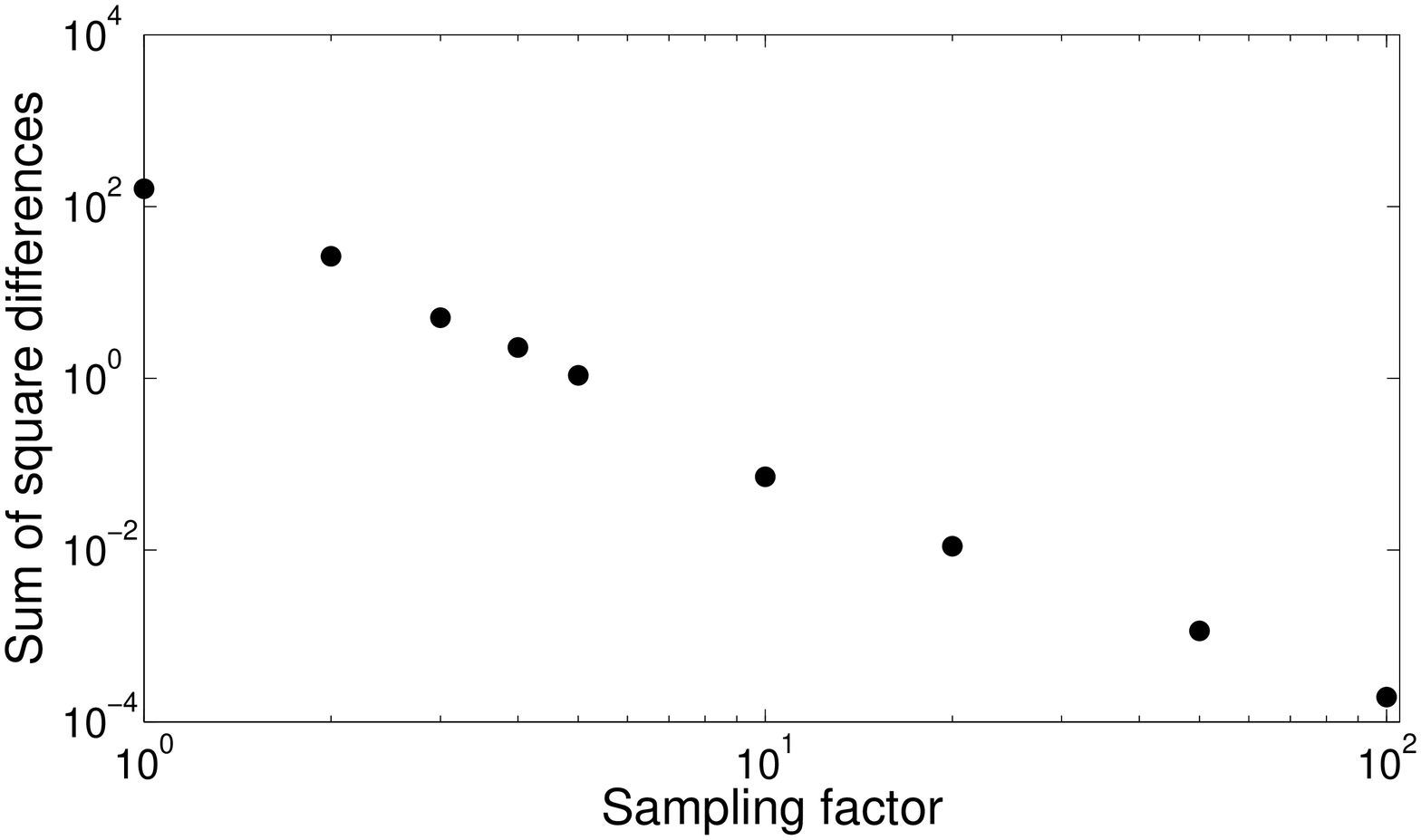}\\
 \caption{
Log-log plot of the sum of square deviations (from the limiting CDF) of the CDF for
$\nu$ times over-sampling vs. the over-sampling rate $\nu$. }
 \label{figure5a}
\end{figure}
The sampling rate sufficient to guarantee convergence to the limiting CDF must be
established individually for each data set. This can be done using plots like those
shown in Figures \ref{figure5} and \ref{figure5a}.

If we are interested in pinpointing precisely the frequency of a peak (as we are in
asteroseismology), then gross over-sampling may be the route to follow. However,
there is a limit to the amount of information contained in the periodogram of a
finite time series. There is therefore also a limit to how finely the frequency axis
should be subdivided. This limit is given by $\Delta \omega_{\rm min} = \pi/T$,
which is the smallest frequency interval that can reasonably be resolved by the data
set. Dense over-sampling in pursuit of the convergence limit of the CDF may lead to
a choice of $\Delta \omega$ smaller than this interval. If the limiting CDF differs
substantially from that obtained from $\Delta \omega_{\rm min}$, then limiting
the sampling interval to $\Delta \omega_{\rm min}$ may be a better option.

As noted previously, \citet{sc98} has little confidence in this method. Apart from
the comments already reported, he further says, p 832, ``Experiments often
demonstrate difficulties in the reproduction of theoretical single-trial
distributions by simulations. Hence the analytical single-trial probabilities
discussed here are essential for the verification of Monte Carlo simulations". He
has a similar comment on p 833 from which he draws the conclusion that
``single-trial analytical probability distributions are indispensable in any
strategy for bandwidth correction". It is not clear how Schwarzenberg-Czerny intends
the phrase ``single-trial probabilities" to be understood, but however we interpret
it, these comments leave us with the same dilemma. It seems to us that our only
recourse in assessing periodogram peak significance in the general case of unevenly
spaced data is Monte Carlo simulation. Though the problems with Monte Carlo
simulations pointed out by Schwarzenberg-Czerny are real, they are not problems of
principle, but of practical implementation. They therefore can, and will, be
overcome in time, if they have not been overcome already.


\section{SUMMARY AND CONCLUSIONS}

Currently available theoretical false alarm probability functions are all derived
from what appear to be reasonable assumptions about the data to be analysed and
about the periodograms that they yield. Their validity, reliability and usefulness
therefore strongly depend on how well these assumptions are met in practice.

A key assumption made by all authors is that the frequency range inspected in the
periodogram contains a set of  $N_i$ frequencies ${\omega_k : k = 1, ..., N_i}$ at
which the periodogram powers ${P_X(\omega_k)}$ are mutually independent. In
theoretical statistical analysis, we have little hope of obtaining a false alarm
probability function in the absence of this assumption. Without independence, very
few general statistical results are available, and none are relevant to the problem
at hand. The assumption of the existence of independent frequencies is therefore
necessary in any theoretical discussion of the problem of significance of
periodogram peaks and poses the first and most important obstruction to its
resolution.

The existence of independent frequencies is guaranteed when the data are evenly
spaced. We should therefore be able to test the validity of proposed false alarm
probability functions for this case against the results of Monte Carlo simulations.
Reasonable requirements on candidate functions include a good fit to the empirical
CDF's, and their ability to predict correctly the number of independent frequencies
known to exist from the theory.

Though they seem not to have viewed their work in this light, HB effectively
performed this test for Scargle's false alarm function. They constructed the
empirical CDF for periodogram peak heights produced by a pure noise time series
consisting of $N_0$ evenly spaced data points, and fitted the Scargle false alarm
function to it by least squares using the number $N_i$ of independent frequencies as
the fitting parameter. According to the theory, they should have obtained $N_i
\lesssim [N_0/2]$. However, their results consistently yielded $N_i > N_0$. Horne
and Baliunas did not comment on this anomaly.

We have repeated their simulations, obtaining results similar to theirs only for
gross over-sampling of the periodogram, and only for data sets with $N_0\leq 170$
data points. For gross over-sampling and data sets with $N_0 > 170$, we were unable
to reproduce their results. The values of $N_i$ obtained by HB are consistently and
systematically larger than ours. In our simulations, the best-fit value of $N_i$
increases linearly with $N_0$, in conflict with the quadratic dependence claimed by
HB. Inspection of a plot of the values published in HB appears to indicate that
their points lie on two straight lines, with a disjunction of slope at $N_0  = 170$
data points. We conjecture from these results that HB constructed their empirical
CDF's by gross over-sampling of the periodogram. This might explain why they
consistently obtained $N_i > N_0$. We also conjecture that the sharp disjunction in
slope at $N_0 = 170$, which is not observed in our simulations, is due to a
systematic error in theirs. If so, the quadratic dependence of $N_i$ on $N_0$,
sometimes exploited by astronomers in the analysis of their data, might not be
a real feature of real astronomical data but rather a spurious artifact of the HB
simulations.

Given the assumption of independence that lies at the heart of Scargle's derivation
of his false alarm probability function, it seemed unreasonable to suppose that it
would provide an adequate description of the empirical CDF's obtained by
over-sampling the periodograms. Accordingly, we initially ran the HB simulations for
CDF's constructed by sampling the periodograms only at the natural frequencies. The
best-fit values of $N_i$ were very close to the theoretically expected value of
$[N_0/2]$. Though heartening, the results of these simulations displayed a
disconcerting feature: the best-fit Scargle functions were very poor fits to the
empirical CDF's, displaying large deviations from the empirical CDF's in the domain
of most interest when assessing the significance of periodogram peaks. The
theoretical false alarm functions were consistently substantially higher in value
than the empirical CDF's, leading to severe under-estimation of peak significance.
This same behaviour was observed for the best-fit curves to the CDF's constructed by
over-sampling the periodograms. Researchers using the Scargle false alarm function,
with or without the HB algorithm, are thus at significant risk of rejecting peaks
that reflect real periodicities in their data.

A flaw in Scargle's derivation of his false alarm function was pointed out by
\citet{koen90} and by \citet{sc96}): Scargle assumes that the variance $\sigma_X^2$
of the noise is known {\em a priori}. This condition is not satisfied either in the
simulations (where we {\em sample} pure pseudo-Gaussian noise), nor in real data
sets (where the data variance must be estimated from the data themselves). Both Koen
and Schwarzenberg-Czerny correct this error in their respective treatments of the
problem. \citet{koen90} concludes that Scargle's false alarm function should be
replaced by the Fisher (or, Fisher-Snedecor) distribution. \citet{sc98} pointed out
that the Fisher distribution is applicable only for ratios of {\em independent}
random variables. With ratios of random variables that are not independent, the
Fisher distribution must be replaced by the incomplete $\beta$-function. He also
showed that, in the case of the Lomb-Scargle periodogram, the correct distribution
is given by the incomplete beta function. On the strength of the work of these
authors, we tested Schwarzenberg-Czerny's proposed function on CDF's constructed by
over-sampling periodograms and also on CDF's constructed by sampling only at the
natural frequencies. In both cases, we have found the best-fit Schwarzenberg-Czerny
function, obtained by the HB algorithm, consistently to fit the empirical CDF's far
more closely than Scargle's function, with impressively good agreement on all but
the smallest data sets, where the theoretical function deviates only slightly from
the empirical CDF's.

In spite of the excellent fits provided by the Schwarzenberg-Czerny false alarm
function, our simulations display an alarming feature: the best-fit values of $N_i$
that yield such excellent agreement with the empirical CDF's are all consistently
higher than the theoretically expected value of $[N_0/2]$. This is not unexpected
for CDF's constructed by over-sampling. In the case of CDF's constructed by sampling
only at the natural frequencies, however, this result is in conflict with the
theory. This means that, as in the case of the Scargle function, we cannot interpret
the best-fit value of the parameter $N_i$ as the number of independent frequencies.
It must be regarded rather as a fitting parameter in a one-parameter family of
candidate CDF functions that fit the empirical CDF's better than Scargle's candidate
functions. Note that the Schwarzenberg-Czerny false alarm functions constructed
independently of simulations, relying exclusively on the use of {\em a priori}
theoretical values for $N_i$ {\em badly overestimate the significance of periodogram
peaks and may result in the acceptance of spurious peaks as genuine}. It would seem
therefore that unqualified confidence in analytical single-trial probability
distributions in the construction of false alarm probability functions may be
misplaced. Even in those cases where they ought to provide a good description of the
behaviour of the empirical CDF's, they apparently fail to do so, leaving us no
option but to resort to Monte Carlo simulations and to treat the theoretical
distributions as nothing more than candidate CDF functions to be accepted or
rejected according to their utility in providing a good fit to the empirical curves
in the region of interest.

Ultimately, our principal interest is in the case of unevenly spaced data, not data
that are evenly spaced. The loss of independence of the variables $P_X(\omega)$ in
this case calls into question the validity and the expediency of searching for a
formula in closed form for a false alarm probability function. All formulae proposed
hitherto are based on the assumption of the existence of a set of mutually
independent periodogram powers. This assumption is not realistic in uneven sampling
schemes, as shown by \citet{koen90}. Were a set of approximately uncorrelated
periodogram powers to be found, in the sense outlined by Scargle, this still would
not guarantee their approximate independence. The currently proposed closed-form
formulae therefore cannot be expected to provide accurate false alarm criteria. Even
if we adopt the attitude that the proposed formulae are no more than candidate CDF
functions, the value of the parameter $N_i$ is not known {\em a priori},
independently of Monte Carlo simulations. Therefore, {\em theoretical probability
distributions provide no predictive power in determining false alarm criteria
appropriate to a given data set which is independent of the empirical CDF's
generated by simulations}. In the final analysis, the only way to obtain the
appropriate false alarm probability function is by first constructing empirical
CDF's for the maximum peak heights by using Monte Carlo methods, and then fitting
these distributions with the false alarm function of choice. If a sufficiently good
fit is obtained, the fitted function can then be used to calculate the significance
levels for the given data set. If the fit is not good however, the significance
levels predicted by these fitted functions are likely to lead to erroneous rejection
or acceptance of periodogram peaks, making them almost useless in the assessment of
the significance of peaks.

At first, this dilemma appears irresolvable. On reflection, however, its resolution
is staring us in the face. What we need is a reliable false alarm probability
function. Though we do not possess this function as a closed-form formula, we
nevertheless have a numerical plot of it in the form of the CDF of maximum peak
heights. This plot can be used just as easily as any closed form formula to get the
answers that we want. If we insist on having a closed-form formula to facilitate
significance estimation, the empirical CDF can be fitted {\em in the region of
interest} by any number of candidate fitting-functions. This renders the need to
search for a theoretical formula obsolete. Of course, it would be nicer, more
convenient and more satisfying to have a theoretical formula, but a numerical plot
of the same function is almost as good.

In this paper we have studied almost exclusively significance tests for the
rejection of spurious peaks in periodograms of data sets that are evenly-spaced in
time. An analogous study for unevenly-spaced data sets is currently in preparation.

\acknowledgments

We sincerely thank Mike Gaylard, Chris Koen and Melvyn Varughese for their critical
reading of draft versions of the manuscript. We also thank the South African SKA
Office in Johannesburg for use of their facilities.

\appendix

\section{CLASSICAL PERIODOGRAM}
\label{classicalP}

The {\em classical periodogram} is essentially a Fourier power spectrum estimator
for an infinite continuous-time signal $X(t)$ that has been discretely sampled for a
finite time at equally spaced time intervals. The data for this estimator form a
finite discrete time series consisting of $N$ values $X_i=X(t_i)$, $i=1,..., N$, of
the physical parameter $X$ at times $t_i= t_0, t_0+\Delta t, t_0+2\Delta t,..., t_0+
(N-1)\Delta t$. The discrete Fourier transform (DFT), $DFT_X (\omega)$, of this time
series $X_i$, which is defined by
\begin{eqnarray}
    DFT_X (\omega) = \sum_{r=1}^{N} \ X(t_r)\, e^{-i\omega t_r}    \label{cp-1}
\end{eqnarray}
may be regarded as an estimator of the Fourier transform $FT_X(t)$ of $X(t)$. The
power spectral density of the signal may then be estimated by the function
\begin{eqnarray*}
    \left| DFT_X (\omega) \right|^2     
\end{eqnarray*}
with some suitably chosen normalising coefficient. A commonly used normalisation is
\begin{eqnarray}
    CP_X (\omega) &=& \frac{1}{N} \, \left| DFT_X (\omega) \right|^2
    = \frac{1}{N} \, \left| \sum_{r=1}^{N} \ X(t_r)\, e^{-i\omega t_r}  \right|^2
       \label{cp-2}
\end{eqnarray}
A simple calculation then yields the formula,
\begin{eqnarray}
    CP_X (\omega) = \frac{1}{N} \, \left[ \left( \sum_{r=i}^N\ X(t_r) \cos \omega t_r \right)^2
    + \left( \sum_{r=i}^N\ X(t_r) \sin \omega t_r \right)^2 \right]     \label{cp-3}
\end{eqnarray}
Following \citet{scargle82}, we call this function the {\em classical periodogram}.
This definition agrees with that given originally by Schuster in \citet{schuster98},
but not with that in his later publications. It also agrees with the definitions
used in \citet{thompson71} and \citet{deeming75}, and differs by a factor of two
from that used by \citet{priestley81}.

It is easy to see from equation (\ref{cp-2}) why the classical periodogram is useful
in identifying the frequencies of harmonic components in the signal $X$. Suppose $X$
contains an harmonic component of frequency $\tilde{\omega}$. Then, when $\omega$ is
very different from $\tilde{\omega}$, $X(t)$ and $e^{-i\omega t}$ are out of phase,
and the product $X(t) e^{-i\omega t}$ oscillates rapidly. The sum of the products
$X(t_r) e^{-i\omega t_r}$, which is a discrete estimator of the integral $\int
\,X(t) e^{-i\omega t}dt$ will thus have a value close to zero, albeit masked by
whatever other signal is present in $X(t)$. As $\omega$ approaches the value of
$\tilde{\omega}$, the factors $X(t)$ and $e^{-i\omega t}$ get closer in phase, so
the product $X(t) e^{-i\omega t}$ oscillates more slowly. The value of the sum of
the products $X(t_r) e^{-i\omega t_r}$ will thus rise, reaching a maximum at $\omega
= \tilde{\omega}$. The presence of a harmonic signal of frequency $\tilde{\omega}$
thus produces a peak in the periodogram with maximum at $\tilde{\omega}$.

The converse however is not true. A peak in the periodogram does not necessarily
reflect the presence of an harmonic component in the signal $X$. Peaks might be
produced by other effects. Thus, the presence of measurement error, signal noise, or
random physical processes in the observed system might, by a spurious random
fluctuation, also produce a peak. Peaks may also be produced by aliasing and / or
spectral leakage, and the observing window. The potential for producing peaks that
are not due to harmonic components in the observed signal makes the interpretation
of peaks in the periodogram very difficult and presents many hazards and pitfalls
for the unwary. The dangers posed by these effects were already noted by Arthur
Schuster as early as 1906, ``... it has generally been assumed that each maximum in
the amplitude of a harmonic term corresponded to a true periodicity. The extent to
which this fallacious reasoning has been made use of would surprise anyone not
familiar with the literature of the subject." (\citet{schuster06}, p 71-72).
Strangely, his warning has often been ignored, and sometimes even disdainfully
brushed aside.

\section{LOMB-SCARGLE PERIODOGRAM}
\label{scargleP}

Following \citet{scargle82}, Appendix B, we define the Lomb-Scargle periodogram by
the formula
\begin{eqnarray}
  P_X (\omega )  &=&  \frac{1}{2} \left\{
  \frac{\left[ \displaystyle \sum_{i=1}^N x_i \cos \omega (t_i - \tau)\right]^2 }
  {\displaystyle \sum_{i=1}^N \cos^2 \omega (t_i - \tau)}
  \ + \ \frac{\left[ \displaystyle \sum_{i=1}^N x_i \sin \omega (t_i - \tau)\right]^2 }
  {\displaystyle \sum_{i=1}^N  \sin^2 \omega (t_i - \tau)} \right\}
  \label{lomb-scargle}
   \end{eqnarray}
where the epoch translation parameter $\tau(\omega)$ is defined implicitly by the
formula
\begin{eqnarray}
  \tan ( 2\omega \tau )  &=&  \frac{ \displaystyle  \sum_{i=1}^N \sin ( 2 \omega t_i )}
  {\displaystyle \sum_{i=1}^N \cos ( 2 \omega t_i )}
  \label{lomb-scargle-A}
   \end{eqnarray}
The data used to calculate $P_X(\omega)$ form a finite discrete time series
consisting of $N$ values $X_i=X(t_i)$, $i=1,..., N$, of the physical parameter $X$
measured at times $\{t_i | i= 1,2,...,N\}$ which are arbitrarily spaced in time.
\citet{lomb76}, following \citet{barning63} and \citet{vanicek69}, arrived at this
formula via a least squares fitting procedure in which sampled values $X(t_i)$ are
fitted with an harmonic signal of frequency $\omega$. For these three authors
therefore, $P_X(\omega)$ does {\em not} represent and attempt at estimating the
Fourier power spectrum of any continuous time physical signal $X(t)$, but is rather
a {\em spectral best-fit parameter} that displays how closely the data may be fitted
with a single harmonic function of frequency $\omega$. The larger the value of
$P_X(\omega)$, the better the fit.

In contrast with these authors, \citet{scargle82} arrived at the same spectral
function by first relaxing the definition of the DFT for application to the case of
unevenly spaced data \citep[Appendix~A]{scargle82}, and then imposing two demands on
the periodogram (which he calls the {\em modified}, or {\em generalised
periodogram}) calculated from this relaxed definition: the statistical distribution
of the generalised periodogram will be made as closely as possible the same as it is
in the evenly spaced case; and, the generalised periodogram will be made invariant
to translations in time. These two requirements yield uniquely the formulae in
equations (\ref{lomb-scargle}) and (\ref{lomb-scargle-A}). Arguably, this restores
the interpretation of the modified periodogram as an estimator of the power spectrum
of the physical signal $X(t)$ in the case where the signal is unevenly sampled in
time. However, it is probably more accurate to regard it as a spectral
goodness-of-fit parameter. This view also enables one better to understand a variety
of other, alternative, periodogram formulae currently offered in the literature.

\section{Pure Noise}
\label{noiseApp}

A random process $X(t)$ is said to be a {\em purely random process}, {\em pure
noise}, or {\em white noise}, if it consists of a sequence of {\em uncorrelated
random variables}. This means that, for all $t'\neq t$,
   $$ {\rm cov}\left(X(t), X(t') \right)=0 $$
Pure noise is the simplest of all random process models. It corresponds to a case
where the process has ``no memory" in the sense that the value of the random
variable $X(t)$ at time $t$ has no relation whatever to its value $X(t')$ at any
other time $t'$, no matter how close or distant $t$ and $t'$ are to each other. In
this sense, $X(t)$ neither remembers its past, nor is aware of its future. Knowing
the value of $X(t_0)$ at any time $t_0$ therefore provides no way at all, other than
by the probability distribution $p_{X(t)}=p(x,t)$, of predicting within reasonable
limits and uncertainties the value of $X(t)$ at time $t$. This is to be contrasted
with {\em correlated noise} where the values $X(t)$ and $X(t')$ are in general
related or `correlated'. In this case, we can do better in predicting the value of
$X(t+\tau)$ from $X(t)$ than in the case of uncorrelated, or pure, noise. From a
knowledge of the value $X(t)$, we can set narrower limits on the probable values of
$X(t+\tau)$  than is possible from the distribution $p_{X(t+\tau)}$ alone. \citep[p
114]{priestley81}.

Pure noise is said to be {\em Gaussian} if the random variables $X(t)$ are jointly
normally distributed. Noise of this kind is often called {\em Gaussian white noise}.
In this case, the random variables $\{X(t)\}$ are also mutually independent.

Note that some authors define pure noise more stringently. For them, a random
process $X(t)$ is pure noise if the random variables $\{X(t)\}$ are {\em
independent}, and {\em identically distributed with zero mean.}

In this paper, a data set $\{ X_k | k=1,2,...,N_0\}$ is said to be pure noise if the
values $X_k$ are {\em independent}, and {\em identically distributed random
variables with zero mean}. For simplicity, we assume also that the $X_k$ are each
normally distributed. Denote their common variance by $\sigma_X^2$. Since the $X_k$
have zero mean, their covariance matrix is given by
\begin{eqnarray}
   C_{jk} = E [ (X_j-\mu_{X_j}) ( X_k -\mu_{X_k})]
   =  E [ X_j X_k ] = \sigma^2_X
  \delta_{jk}
\end{eqnarray}


\end{document}